\documentclass[dvips,traditabstract]{aa}

\usepackage{graphicx,amssymb,amsmath,hyperref,multirow}
\usepackage[usenames,dvipsnames]{xcolor}
\usepackage{adjustbox}

\usepackage{rotating}
\usepackage{txfonts}
\usepackage{lscape}
\usepackage{amsmath}
\usepackage{natbib,twoopt}
\usepackage{longtable}
\usepackage{supertabular,booktabs}
\usepackage{tabularx}
\bibpunct{(}{)}{;}{a}{}{,} 
\newcommandtwoopt{\citeyearads}[3][][]%
{\href{http://adsabs.harvard.edu/abs/#3}{\citeyear[#1][#2]{#3}}}
\hypersetup{colorlinks,linkcolor=red,citecolor=blue,urlcolor=blue}

\newcommand{\feh}{[Fe/H]}
\newcommand{\teff}{$\mathrm{T}_{\mathrm{eff}}$}
\newcommand{\logg}{$\log g$}
\newcommand{\vmic}{$v_{\mathrm{mic}}$}

\newcommand{\tab}[1]{Table~\ref{#1}}
\newcommand{\fig}[1]{Fig.~\ref{#1}}
\newcommand{\sect}[1]{Sect.~\ref{#1}}

\newcommand{\hd}{HD~142575}
\newcommand{\cs}{CS~22874$-$042}
\newcommand{\cd}{CD~$-$48~2445}

\begin{document}

\title{Something borrowed, something blue: The nature of blue metal-poor stars inferred from their colours and chemical abundances
\thanks{Based on UVES archive data 077.B-0507 and 090.B-0605.
This paper includes data gathered with the 6.5 meter Magellan Telescopes located at Las Campanas Observatory, Chile.
%
}
}
\author{
	C.~J. Hansen \inst{\ref{dark},\ref{lanc}}, 	
	P. Jofr\'e \inst{\ref{ioa}, \ref{udp}},
	A. Koch\inst{\ref{lanc}},
	A. McWilliam \inst{\ref{carnegie}},
	\and C.~S. Sneden \inst{\ref{texas}}
	}

\authorrunning{Hansen et al. }
\titlerunning{Lithium in BMP stars}
\offprints{ \\ 
C. J. Hansen, \email{cjhansen@dark-cosmology.dk}}

\institute{Dark Cosmology Centre, The Niels Bohr Institute, Juliane Maries Vej 30, DK-2100 Copenhagen, Denmark \label{dark}
	\and Department of Physics, Lancaster University, Bailrigg, Lancaster LA1 4YB, UK \label{lanc}
	\and Institute of Astronomy, University of Cambridge, Madingley Road, Cambridge CB3 0HA, United Kingdom \label{ioa}
	\and {N\'ucleo de Astronom\'ia, Facultad de Ingenier\'ia, Universidad Diego Portales,  Av. Ej\'ercito 441, Santiago, Chile}\label{udp}
	\and  Carnegie Observatories, 813 Santa Barbara St., Pasadena, CA 91101, USA \label{carnegie}
	\and Department of Astronomy, University of Texas at Austin, 1 University Station C1400, Austin, TX 78712, USA \label{texas}
	}

   \date{}

\abstract{Blue metal-poor stars (BMPs) are { main sequence stars that appear bluer and more luminous than normal turnoff stars. They were originally singled out by using $B-V$ and $U-B$ colour cuts. }
Early studies found that a larger fraction of field BMP stars were binaries compared to normal halo stars. Thus, BMP stars are ideal field blue straggler candidates for investigating internal stellar evolution processes and binary interaction.
In particular, the presence or depletion in lithium in their spectra is a powerful indicator as to their origin. They are either old, halo blue stragglers experiencing internal mixing processes or mass transfer (Li-depletion), or intermediate-age, single stars of possibly extragalactic origin (2.2\,dex halo plateau Li). However, we note that internal mixing processes can lead to an increased level of Li. Hence, this study combines photometry and spectroscopy to unveil the origin of various BMP stars. We first show how to separate binaries from young blue stars using photometry, metallicity, and lithium. Using a sample of { 80} BMP stars (T$> 6300$K), we find that 97\% of the BMP binaries have $V-Ks_0 < 1.08\pm0.03$, while BMP stars that are not binaries lie above this cut in 2/3 of the cases.  This cut can help classify stars which lack radial velocities from follow-up observations. 
We then trace the origin of two BMP stars from the photometric sample by conducting a full chemical analysis using new high-resolution and high signal-to-noise spectra. Based on their radial velocities, Li, $\alpha$, and s- and r-process abundances we show that BPS CS22874-042 is a single star (A(Li)$=2.38\pm0.10$\,dex) while with A(Li)$=2.23\pm0.07$\,dex CD-48 2445 is a binary, contrary to earlier findings.
Our analysis emphasises that field blue stragglers can be segregated from single metal-poor stars, using ($V-Ks$)  colours with a fraction of single stars polluting the binary sample, but not vice versa. These two groups can only be properly separated by using information from stellar spectra, illustrating the need for accurate and precise stellar parameters and high resolution, high S/N spectra in order to fully understand  and classify this intriguing class of stars. { Our high-resolution spectrum analysis confirms the findings from the colour cuts and shows that \cs\ is single, while \cd\ most likely is a binary. Moreover, the stellar abundances show that both stars formed in situ; \cs\ carry traces of massive star enrichment and \cd\ shows indications of AGB mass transfer mixed with gases ejected possibly from neutron star mergers.}
}
\maketitle
 
 \section{Introduction}
Amongst the many intriguing facets of the Galactic halo, the near-constancy 
in lithium abundance, at A(Li)=2.2\,dex over a broad range in temperature and metallicity (the so-called ``Spite-plateau''), is now a well-established
property of old and warm halo main-sequence stars \citep{Spite1982}. 
Thus, a star presenting lithium abundances that deviate from the plateau value, either through strong enhancements or depletions,
suggests that it has experienced internal or external alterations that are not present 
in its halo star counterparts, motivating us to investigate these physical processes that have changed its lithium abundance. 

Metal-poor dwarfs depleted in lithium are relatively common  \citep[e.g.,][]{Sbordone2010} and one option is  that they have suffered from mass transfer or merging between stars  \citep[e.g.,][]{Carney2005}. On the other hand, 
there are far fewer cases of metal-poor dwarfs which have Li abundances significantly 
higher than the plateau value \citep{Deliyannis2002,Asplund2006, Koch2011, Monaco2012}, and it is still not clear how these form.  
Possible explanations include diffusion within the star \citep{Richard2005},  
seeding by supernovae \citep[SNe;][]{Woosley1995}, asymptotic giant branch (AGB) companions \citep{Ventura2011}, 
or accretion of substellar bodies \citep{Ashwell2005}. 

{ Blue stragglers are excellent candidates to learn about the processes that can create and destroy lithium in main-sequence stars. These stars can obtain their blue colour and low Li in a number of ways; through mass transfer, coalesence, pulsational driven mass loss or internal mixing, where mass transfer seems to be the preferred scenario \citep{ps00}. Hence, blue stragglers are bluer than turn-off stars of a coeval population because they became more massive thanks to direct mass transfer from a companion star in a binary system. If mass transfer has taken place a higher rotation velocity will often testify to this, and Li will subsequently be destroyed. The binary system will typically consist of two low-mass stars, where one of them is a white dwarf ($<0.55$M$_{\odot}$ on an almost circular orbit with a period of a few hundred days \citep{Carney2005}).
Thus, spectroscopic studies of blue stragglers in the halo help to investigate processes of mass transfer, which Li excellently traces \citep{Carney2005}. In this study, we therefore present analyses of photometry and spectroscopy, as both are needed to pin down the true origin of the BMP stars. 
}

Contrary to globular clusters, field blue stragglers are difficult to find because of the variety of metallicities and distances that stars in the Galactic halo cover. 
Although the inner halo is composed of one dominant co-eval population, presenting a well-defined turn-off colour as a function of metallicity \citep[e.g.,][]{1991ApJS...76.1001P, 1996MNRAS.278..727U, 2011A&A...533A..59J}, not all main-sequence stars that are bluer than this turn-off temperature are necessarily blue stragglers. They could also be intermediate-age stars with extragalactic origin that were accreted later onto the Milky Way \citep{1991ApJS...76.1001P, 1996MNRAS.278..727U}, or simply warm, metal-poor single stars formed in situ.   
In this context, the radial velocity variations and the abundance of lithium give key information about the nature of BMP stars: while blue stragglers should present overall, strong modifications of their Li-content, 
intermediate-age stars should be Li-normal \citep{Ryan2001}. 
To determine if the star has an extra-galactic origin, the $\alpha-$abundances are useful discriminants \cite[see, e.g.,][]{Nissen2010} and \citealt{2016A&A...590A..39H}).

The largest sample of BMP stars to date is still the one published 16 years ago by \citet[][hereafter PS00]{ps00}. The stars were selected using a $B-V$ vs. $U-B$-diagram.
PS00 analysed a sample of 62 BMP stars using high-resolution, { low signal-to-noise ratio (SNR$\sim10-30)$} spectroscopy and monitored their radial velocities (RVs) over { seven} 
years. 
From this,  \citet{2003ApJ...592..504S} found a systematic difference in the $s-$process enhancement between stars with constant and varying radial velocities (RV).  
The study of PS00  also revealed  that 17 stars  had constant  RV and more than 50\% of them were { low} in $\alpha-$elements, and thus good contenders for intermediate-age, accreted stars.  The Li abundances in these intermediate-age candidates have not been studied so far.  

It is important to note that the Li resonance line at 6707.8\,\AA\ becomes very weak in BMP stars, where an abundance at the  
Spite-Plateau value of A(Li)=2.2 produces an equivalent width (EW) of less than 10\,m\AA\ \citep{2015ebss.book...65P}.  
Furthermore, one needs to be aware that a lithium abundance alone is not a clear discriminator between (Li depleted)  blue stragglers and  
intermediate-age, metal-poor dwarfs with (generally) a normal Li abundance.  
For instance, 
in the Hyades, dwarfs with similar temperatures to the BMP stars in PS00 are depleted in Li, producing the so-called ``Li-dip" \citep{Boesgaard1986}.   
\citet{Carney2005} addressed this issue and showed that one BMP star, at constant RV in the temperature range of the Li-dip (between 6400 and 6800~K) is depleted in Li  by $\sim$0.7\,dex while  another star
had normal plateau value. These authors suggested that, at low metallicities, the temperature range of the Li-dip  could be shifted towards hotter temperatures, or that no Li-dip exists at low metallicities. However, \citet{Asplund2006} and \citet{Bonifacio2007}
found a first indication for this feature among metal-poor Population II stars, which 
they relate to effects of mass loss on the main sequence, as predicted by \citet{Dearborn1992}. 

{ In this work, we aim to gain insights into to possible mechanisms that can alter the Li abundances in field blue stragglers or BMP stars. Therefore, we need to perform not only a Li abundance analysis of the BMP stars but also a detailed spectral analysis of $\alpha-$, r- and s-process elements. The former two allow us to investigate a Galactic vs. extragalactic origin and exact formation site, while s-element abundances provide information on the presence and amount of mass transfer from a binary AGB companion star. The latter scenario can also be tested through possible RV variations and change in colour. The PS00 sample is an obvious starting point as they provide information on radial velocities, $\alpha-$ and s-process abundances. Unfortunately, their spectra were of too low signal-to-noise to confidently allow for Li and r-element detections in these warm, dwarf stars. This prompted the need for high SNR spectra.} 
\begin{figure}
\includegraphics[width=1\hsize]{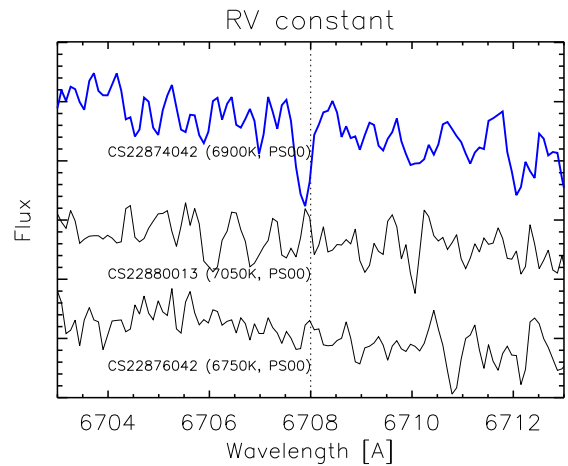}
\centering
\caption{Spectrum taken from PS00 of the star \cs\ around the Li line at 6707.8\,\AA. Stars with similar temperatures (according to PS00) shown for comparison.}
\label{li_ps00}
\end{figure}

{ The RV-constant star, CS~22874-024, turned out to be an outlier in PS00 as it 
showed a  notable Li line of $\sim26~\mathrm{m\AA}$ (see \fig{li_ps00}). }
According to the stellar parameters of CS~22874-024 reported in PS00, this would correspond to an abundance of A(Li) = 3.25,  which is 
significantly higher than the Spite-plateau value, as well as the value expected from Big Bang Nucleosynthesis \citep[BBN;][]{Cyburt2008}. { High resolution, high SNR spectra are clearly needed to confirm the stellar parameters and accurately measure key tracer abundances from weaker lines.}

We compile a large sample of { 80} BMP stars from the literature to devise a method to separate blue stragglers from metal-poor single stars based on various properties. In particular, we show that 
their Li and heavy element abundances plus their colours are imperative to 
segregate the BMP stars in subsamples (\S2). 
In \S3 we introduce a subsample of stars including BMP and benchmark stars, for which we 
derive detailed chemical abundances. 
The results from our spectral analysis are presented in  \S4, which we 
discuss in \S5 before  concluding our results in \S6. 

\section{Photometry of the BMP sample}

Here we select warm stars (T$> 6300$K) with gravities around or above 4.0\,dex spanning a broad range of metallicities for a sample consisting of both binaries and single stars with known Li abundances to develop a way of separating single and binary blue stars. { This temperature range is selected so that a possible Li gap may be detected if existing for these field BMP stars.} 
\begin{figure}
\hspace{-1.0cm}
\includegraphics[scale=0.55]{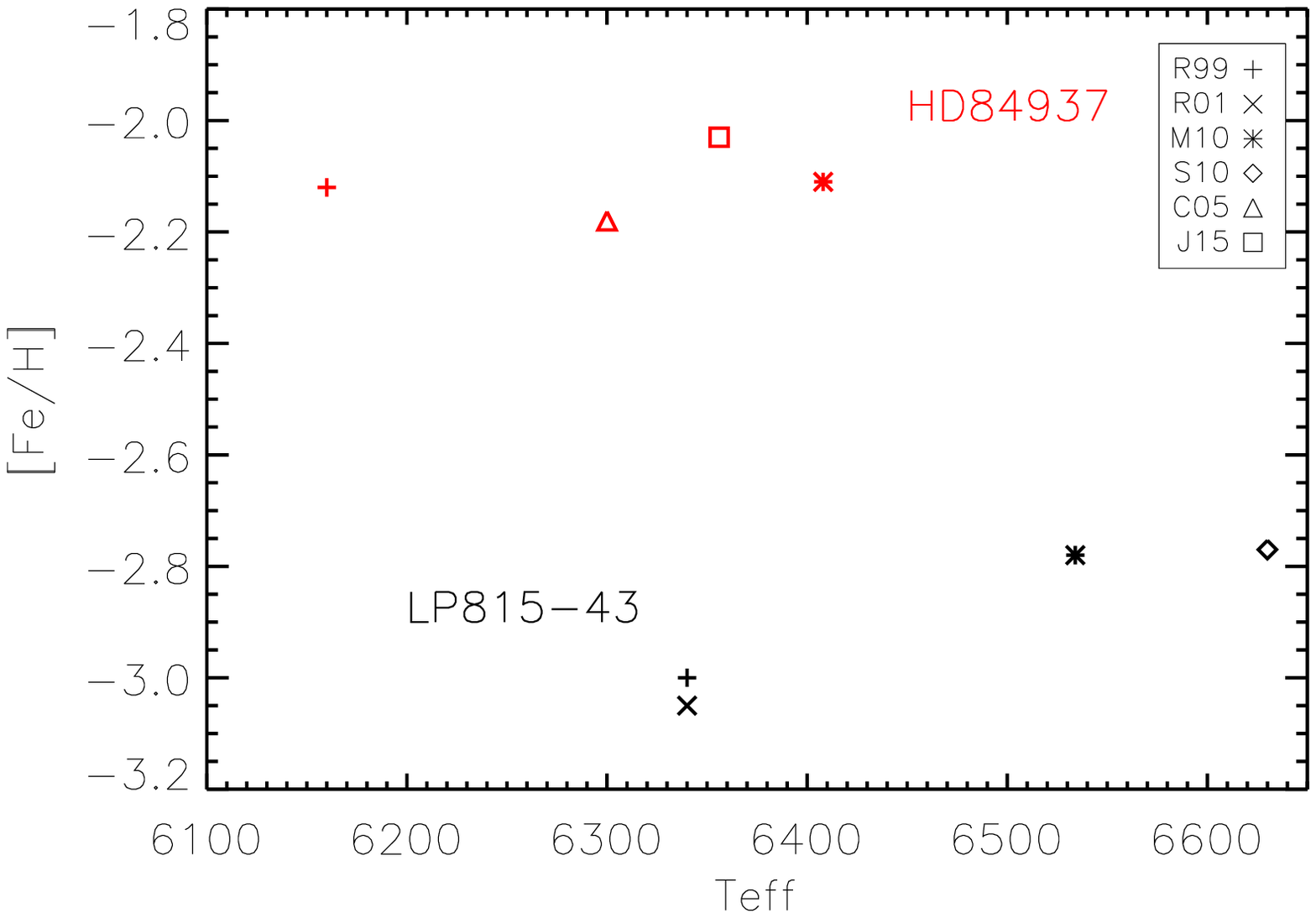}
\centering
\caption{ Temperature and [Fe/H] for two well-studied stars HD84937 and LP815-43
\citet[][R99]{Ryan1999}, \citet[][R01]{Ryan2001}, \citet[][C05]{Carney2005}, \citet[][M10]{Melendez2010}, \citet[][S10]{Sbordone2010}, and
\citet[][J15]{2015A&A...582A..81J}.}
\label{Lit_spread}
\end{figure}

\begin{table*}
 \caption{Coordinates, photometry (from Simbad and literature), dereddening, temperature, metallicity, Li abundance,  radial velocity (RV), and literature reference for the binary and candidate binary BMP sample. An '*' indicates that the star was analysed spectroscopically in this study. Our values are listed in Tables~\ref{bmp_obs}, \ref{Tab:compsample}, and \ref{tab:abun}.}
 \label{tab:photometry1}
\vspace{-0.2cm}
\centering
\resizebox{\textwidth}{!}{\begin{tabular}{l  c c c c c c c c c c}
\hline
\hline
Star	&   Ra($\alpha$), Dec($\delta$) &  V    &   Lit. V     & 	K$_s$  &  E(B-V) &    T  &  [Fe/H]  &     A(Li) & 	RV  & Ref. \\ 
& (J2000.0) & [mag] &  [mag] & [mag] & [mag] & [K] & dex &  dex & [km/s] & \\
\hline
Binaries &&&&&&&&\\
BD+51 1817  	&  13 08 39.10, +51 03 59.26      &  10.21   &  10.23   &   9.08   &	      0.000 &	     6345 &	 $-$1.10 &    1.64   &	...  & R01 \\ 
BD+25 1981  	&  08 44 24.69, +24 47 47.75      &  9.29    &  9.29    &   8.39   &	      0.000 &	     6780 &	 $-$1.30 &    1.75   &	...  & R01 \\ 
BD+23 74   	&  00 32 43.32, +24 13 21.41      &  9.88    &  9.88    &   9.11   &	      0.022 &	     7500 &	 $-$0.91 &    1.32   &	32.5  & C05 \\ 
G202$-$65  	&  16 35 58.58, +45 51 59.26      &  11.06   &  11.22   &   10.18  &	      0.000 &	     6390 &	 $-$1.50 &    1.67   &	...  & R01 \\ 
HD8554 		&  01 24 42.30, +07 00 05.23      &  9.57    &  9.57    &   8.63   &	      0.030 &	     6780 &	 $-$1.47 &    1.11   &	11.1  & C05 \\ 
HD109443  	&  12 34 46.73, $-$23 28 32.20	 &  9.25    &  9.25    &   8.16   &	      0.077 &	     6650 &	 $-$0.55 &    0.67   &	43.4  & C05 \\ 
HD135449  	&  15 16 10.39, $-$32 53 33.03      &  9.46    &  9.46    &   8.15   &	      0.285 &	     6740 &	 $-$0.92 &    1.10   &   $-$42.0  & C05 \\ 
    CS22956$-$028 &  21 44 48.72, $-$63 22 09.80	 &  13.00   &  13.00   &   11.98  &	      0.028 &	     7035 &	 $-$1.89 &    2.00   &	...  & L05, PS00 \\ 
    CS29497$-$030 &  00 40 47.93, $-$24 07 33.91	 &  12.66   &  12.70   &   11.75  &	      0.014 &	     6650 &	 $-$2.70 &    1.10   &	...  & S04, PS00 \\ 
    CS22873$-$139 &  20 05 55.15, $-$59 17 11.40	 &  13.83   &  13.80   &   12.63  &	      0.030 &	     6400 &	 $-$2.90 &    2.15   &	...  & R01 \\ 
    CS22166$-$004 &  00 52 11.44, $-$11 04 40.06	 &  13.08   &  13.10   &   12.53  &	      0.028 &	     .... &	 $-$1.30 &   ....   &	...  &  PS00\\ 
    CS22166$-$041 &  01 10 56.36, $-$13 42 43.50	 &  14.36   &  14.30   &   13.34  &	      0.020 &	     7050 &	 $-$1.32 &   ....   &	...  & PS00 \\ 
    CS22170$-$028 &  00 51 07.37, $-$11 08 31.95	 &  11.52   &  11.80   &   11.17  &	      0.027 &	     8050 &	 $-$0.68 &   ....   &	...  & PS00 \\ 
    CS22174$-$040 &  01 28 40.52, $-$07 24 04.30	 &  13.18   &  13.20   &   12.62  &	      0.026 &	     .... &	 $-$1.60 &   ....   &	...  & PS00 \\ 
    CS22872$-$062 &  16 26 37.03, $-$04 32 54.06	 &  14.05   &  14.30   &   12.86  &	      0.200 &	     7900 &	  0.24 &   ....   &	...  & PS00 \\ 
    CS22874$-$034 &  14 38 26.84, $-$23 14 48.49	 &  14.49   &  14.60   &   13.66  &	      0.095 &	     .... &	 $-$1.70 &   ....   &	...  & PS00 \\ 
    CS22876$-$008 &  23 55 21.37, $-$34 48 01.30	 &  13.95   &  13.90   &   13.12  &	      0.012 &	     7200 &	 $-$1.88 &   ....   &	...  & PS00 \\ 
    CS22876$-$021 &  00 01 03.78, $-$33 48 15.76	 &  14.48   &  14.50   &   13.48  &	      0.011 &	     7250 &	 $-$1.10 &   ....   &	...  & PS00 \\ 
    CS22880$-$073 &  20 48 10.95, $-$20 59 23.90	 &  14.05   &  14.10   &   13.19  &	      0.058 &	     .... &	 $-$0.40 &   ....   &	...  & PS00 \\ 
    CS22885$-$048 &  20 19 53.78, $-$38 56 36.07	 &  14.27   &  14.30   &   13.21  &	      0.050 &	     7050 &	 $-$1.36 &   ....   &	...  & PS00 \\ 
    CS22890$-$069 &  15 24 12.28, +03 05 06.43	 &  12.71   &  12.70   &   12.03  &	      0.039 &	     .... &	 $-$2.00 &    0.00  &	...  & PS00 \\ 
    CS22894$-$029 &  23 42 14.62, $-$00 51 32.71	 &  14.37   &  14.40   &   13.48  &	      0.023 &	     7350 &	 $-$1.47 &   ....   &	...  & PS00 \\ 
    CS22896$-$149 &  19 43 44.92, $-$56 10 08.76	 &  13.03   &  12.10   &   11.47  &	      0.040 &	     7950 &	 $-$0.16 &    0.00  &	...  & PS00 \\ 
    CS22941$-$005 &  23 29 52.83, $-$35 13 03.83	 &  14.61   &  14.60   &   13.77  &	      0.016 &	     7450 &	 $-$2.43 &   ....   &	...  & PS00 \\ 
    CS22946$-$011 &  01 19 11.68, $-$19 23 58.12	 &  13.95   &  14.00   &   12.87  &	      0.014 &	     6850 &	 $-$2.59 &   ....   &	...  & PS00 \\ 
    CS22948$-$068 &  21 49 06.93, $-$37 27 52.49	 &  13.77   &  13.70   &   12.78  &	      0.022 &	     7200 &	 $-$1.37 &   ....   &	...  & PS00 \\ 
    CS22963$-$013 &  02 53 53.75, $-$06 42 17.84	 &  13.51   &  13.50   &   12.61  &	      0.045 &	     .... &	 $-$2.50 &   ....   &	...  & PS00 \\ 
    CS22966$-$037 &  23 43 38.60, $-$31 48 18.40	 &  14.02   &  14.00   &   13.28  &	      0.013 &	     7400 &	 $-$1.34 &   ....   &	...  & PS00 \\ 
    CS22966$-$043 &  23 43 54.45, $-$28 18 34.50	 &  13.56   &  13.60   &   12.83  &	      0.014 &	     7300 &	 $-$1.96 &   ....   &	...  & PS00 \\ 
    CS22966$-$054 &  23 49 51.58, $-$29 08 21.45	 &  14.43   &  14.40   &   13.65  &	      0.019 &	     7250 &	 $-$1.17 &   ....   &	...  & PS00 \\ 
    CS29509$-$027 &  00 50 15.78, $-$30 59 56.16	 &  12.72   &  12.50   &   11.47  &	      0.019 &	     7050 &	 $-$2.01 &   ....   &	...  & PS00 \\ 
    CS29518$-$024 &  01 17 12.12, $-$32 26 57.79	 &  14.52   &  14.50   &   13.96  &	      0.020 &	     7650 &	 $-$0.94 &   ....   &	...  & PS00 \\ 
    CS29518$-$039 &  01 22 45.58, $-$28 24 29.84	 &  14.26   &  14.20   &   13.35  &	      0.014 &	     7050 &	 $-$2.49 &   ....   &	...  & PS00 \\ 
    CS29527$-$038 &  00 37 44.69, $-$20 08 11.12	 &  14.41   &  14.40   &   13.60  &	      0.017 &	     .... &	 $-$1.50 &   ....   &	...  & PS00 \\ 
    CS29527$-$045 &  00 35 50.90, $-$17 57 00.57	 &  14.04   &  14.00   &   13.28  &	      0.019 &	     7250 &	 $-$2.14 &   ....   &	...  & PS00 \\ 
\hline
Maybe binaries &&&&&&&&\\
BD+24 1676  	&  07 30 41.26, +24 05 10.25      &  10.80   &  10.80   &   9.54   &	      0.013 &	     6387 &	 $-$2.54 &    2.27   &   $-$238.6  & R99,M10 \\ 
CD$-$24 17504  	&  23 07 20.23, $-$23 52 35.60	 &  12.18   &  12.18   &   10.81  &	      0.020 &	     6451 &	 $-$3.34 &    2.18   &	136.6  & M10 \\ 
CD$-$33 01173  	&  03 19 35.32, $-$32 50 43.14      &  10.90   &  10.94   &   9.75   &	      0.005 &	     6536 &	 $-$3.01 &    2.18   &	47.6  &  R99,M10\\ 
CD$-$48 2445*  	&  06 41 26.66, $-$48 13 14.95	 &  10.54   &  10.54   &   9.29   &	      0.015 &	     6453 &	 $-$1.93 &    2.22,2.38   &	319.2  & M10,A06 \\ 
G064$-$12  	&  13 40 02.50, $-$00 02 18.80	 &  11.45   &  11.45   &   10.21  &	      0.003 &	     6463 &	 $-$3.26 &    2.32   &	442.5  & R99,M10 \\ 
HD338529  	&  19 32 31.91, +26 23 26.14      &  9.37    &  9.37    &   8.14   &	      0.060 &	     6335 &	 $-$2.26 &    2.25   &   $-$128.3  &  A06\\ 
LP831$-$070  	&  03 06 05.43, $-$22 19 17.94      &  11.80   &  11.80   &   10.34  &	      0.005 &	     6414 &	 $-$2.94 &    2.28   &   $-$48.2	& R99,M10 \\ 
    BS16023$-$46 	&  14 00 54.47, +22 46 42.34	 &  14.23   &  14.23   &   12.96  &	      0.004 &	     6547 &	 $-$2.90 &    2.27   &	 ...  &  M10,B07,S10\\ 
    BS17570$-$63 	&  00 20 36.19, +23 47 37.70	 &  14.51   &  14.51   &   13.07  &	      0.026 &	     6318 &	 $-$2.91 &    2.06   &	 ...  & M10,B07,S10 \\ 
    BS17572$-$100 &  09 28 55.35, $-$05 21 40.38	 &  12.19   &  12.19   &   10.95  &	      0.016 &	     6596 &	 $-$2.66 &    2.28   &	191.0/189.0  & M10,B07,S10 \\ 
    CS22177$-$009 &  04 07 40.64, $-$25 02 43.94	 &  14.27   &  14.27   &   12.95  &	      0.023 &	     6421 &	 $-$3.04 &    2.28   &   $-$1.9  &  M10,B07,S10\\ 
    CS22888$-$031 &  23 11 32.47, $-$35 26 42.90	 &  14.90   &  14.90   &   13.58  &	      0.007 &	     6335 &	 $-$3.24 &    2.11   &   $-$126.0  & M10,B07,S10 \\ 
    CS22953$-$037 &  01 25 06.65, $-$59 16 00.70	 &  13.64   &  13.64   &   12.46  &	      0.008 &	     6532 &	 $-$2.84 &    2.27   &   $-$153.0  & M10,B07,S10 \\ 
    CS22966$-$011 &  23 35 07.05, $-$30 22 54.35	 &  14.55   &  14.55   &   13.28  &	      0.000 &	     6307 &	 $-$3.06 &    1.92   &	 ...  & M10,B07,S10 \\ 
    CS29518$-$043 &  01 18 38.30, $-$30 41 02.66	 &  14.57   &  14.57   &   13.37  &	      0.008 &	     6517 &	 $-$3.17 &    2.20   &	 ...  & M10,B07,S10 \\ 
    CS29527$-$015 &  00 29 10.68, $-$19 10 07.25	 &  14.24   &  14.24   &   13.05  &	      0.014 &	     6541 &	 $-$3.43 &    2.25   &	48.0  & M10,B07,S10 \\ 
    CS31061$-$032 &  02 38 43.29, +03 19 02.44	 &  13.90   &  13.90   &   12.61  &	      0.015 &	     6433 &	 $-$2.57 &    2.23   &	...  &  M10,B07,S10\\ 
    CS22174$-$008 &  01 11 45.47, $-$10 40 03.81	 &  13.62   &  13.60   &   12.69  &	      0.025 &	     7320 &	 $-$1.48 &    ....   &	...  & PS00 \\ 
    CS22896$-$103 &  19 33 15.47, $-$54 58 52.01	 &  14.54   &  14.50   &   13.66  &	      0.046 &	     7400 &	 $-$0.10 &    ....   &	...  & PS00 \\ 
    CS29497$-$017 &  00 31 25.75, $-$23 57 26.82	 &  14.14   &  14.10   &   13.44  &	      0.014 &	     7500 &	 $-$1.19 &    ....   &	...  & PS00 \\ 
    CS29499$-$057 &  23 51 32.25, $-$25 45 46.53	 &  13.85   &  13.80   &   13.22  &	      0.019 &	     7700 &	 $-$2.33 &    ....   &	...  & PS00 \\ 
    CS29517$-$021 &  23 58 09.09, $-$14 51 47.30	 &  13.60   &  13.60   &   12.54  &	      0.027 &	     7000 &	 $-$0.99 &    ....   &	...  & PS00 \\ 
\hline
\hline
\end{tabular}}
 \tablefoot{References: R99: \citet{Ryan1999},   PS00: \citet{ps00}, 	R01:, \citet{Ryan2001},					 S04: \citet{Sivarani2004},				
C05: \citet{Carney2005},
L05: \citet{Lucatello2005},
A06: \citet{Asplund2006}, 
B07: \citet{Bonifacio2007},
M10: \citet{Melendez2010},			
S10: \citet{Sbordone2010}. }
\end{table*}
\begin{table*}
 \caption{Coordinates, photometry (from Simbad and literature), dereddening, temperature, metallicity, Li abundance,  radial velocity (RV), and literature reference for the single star BMP sample.  An '*' indicates that the star was analysed spectroscopically in this study. Our values are listed in Tables~\ref{bmp_obs}, \ref{Tab:compsample}, and \ref{tab:abun}.}
 \label{tab:photometry2}
\vspace{-0.2cm}
\centering
\resizebox{\textwidth}{!}{\begin{tabular}{l c c c c c c c  c  c c}
\hline
\hline
Star	&   Ra($\alpha$), Dec($\delta$) &  V    &   Lit. V     & 	K$_s$  &  E(B-V) &    T  &  [Fe/H]  &     A(Li) & 	RV  & Ref. \\ 
& (J2000.0) & [mag] &  [mag] & [mag] & [mag] & [K] & dex &  dex & [km/s] & \\
\hline
Single stars &&&&&&&&\\
BD+03 0740  	&  04 55 43.46, +31 09 00.93	 &  10.76   &  9.80    &   9.51   &	      0.022 &	     6419 &	 $-$2.71 &    2.21   &	174.2  &  R99,A06,M10\\ 
BD+09 2190 	&  09 29 15.56, +08 38 00.46      &  11.14   &  11.15   &   9.91   &	      0.015 &	     6392 &	 $-$2.66 &    2.13   &	266.6  &  R99,A06 \\ 
BD$-$13 3442  	&  11 46 50.65, $-$14 06 43.45      &  10.27   &  10.29   &   9.02   &	      0.011 &	     6311 &	 $-$2.71 &    2.16   &	116.1  &  R99,A06 \\ 
CD$-$3514849  	&  21 33 49.75, $-$35 26 14.23      &  10.63   &  10.57   &   9.29   &	      0.002 &	     6396 &	 $-$2.35 &    2.37   &	108.0  &  R99,A06,M10\\ 
G064$-$37  	&  14 02 30.09, $-$05 39 05.20      &  11.14   &  11.14   &   9.92   &	      0.012 &	     6583 &	 $-$3.17 &    2.21   &	81.5   &  R99,M10\\ 
HD84937  	&  09 48 56.10, +13 44 39.32      &  8.32    &  8.32    &   7.06   &	      0.005 &	     6408 &	 $-$2.11 &    2.32   &   $-$15.0   &  M10,C05\\ 
HD142575  	&  15 55 02.84, +05 04 12.15      &  8.62    &  8.62    &   7.51   &	      0.056 &	     6550 &	$-$0.97 &    1.45   &   $-$65.0   &  C05 \\ 
LP815$-$43  	&  20 38 13.30, $-$20 26 10.87	 &  10.72   &  10.91   &   9.65   &	      0.024 &	     6534 &	 $-$2.78 &    2.26   &   $-$3.6    &  S10,M10,A06\\ 
    CS22874$-$042* &  14 38 01.94, $-$24 58 44.32	 &  13.91   &  13.91   &   12.42  &	      0.076 &	     6500/6900 &	 $-$1.90 &    2.38	&    176.0  & Our,PS00 \\ 
    CS22950$-$173 &  20 35 31.26, $-$15 53 30.60	 &  14.04   &  14.04   &   12.66  &	      0.046 &	     6353 &	 $-$2.73 &    2.21   &	69.0   &  S10,PS00\\ 
    CS22964$-$214 &  20 05 50.00, $-$39 27 43.30	 &  13.66   &  13.70   &   12.32  &	      0.076 &	     6340 &	 $-$3.30 &    2.15   &	...    &  R01,PS00\\ 
    CS22175$-$034 &  02 20 21.52, $-$10 38 08.95	 &  12.60   &  12.60   &   11.62  &	      0.024 &	     7100 &	$-$0.28 &   ....   &	...  &   PS00\\ 
    CS22185$-$009 &  03 14 53.75, $-$14 43 53.71	 &  13.79   &  13.80   &   12.93  &	      0.043 &	     7100 &	 $-$1.67 &   ....   &	...  &  PS00 \\ 
    CS22871$-$040 &  14 39 20.00, $-$20 50 31.62	 &  12.72   &  12.80   &   11.99  &	      0.087 &	     7880 &	 $-$1.66 &   ....   &	...  &  PS00 \\ 
    CS22874$-$009 &  14 34 23.17, $-$26 17 37.62	 &  13.58   &  13.70   &   12.53  &	      0.077 &	     7600 &	$-$0.42 &   ....   &	...  &  PS00 \\ 
    CS22876$-$042 &  00 12 00.83, $-$33 59 50.50	 &  13.12   &  13.10   &   11.90  &	      0.011 &	     6750 &	 $-$2.06 &   ....   &	...  &  PS00 \\ 
    CS22880$-$058 &  20 43 55.98, $-$21 36 32.50	 &  14.54   &  14.60   &   13.25  &	      0.043 &	     7150 &	 $-$1.85 &   ....   &	...  &  PS00 \\ 
    CS22941$-$012 &  23 28 29.73, $-$32 48 46.40	 &  12.45   &  12.50   &   11.47  &	      0.013 &	     7200 &	 $-$2.03 &   ....   &	...  &  PS00 \\ 
    CS22948$-$079 &  21 47 29.21, $-$39 25 19.26	 &  13.72   &  13.70   &   12.62  &	      0.024 &	     6700 &	 $-$1.63 &   ....   &	...  &  PS00 \\ 
    CS22950$-$078 &  20 24 57.80, $-$16 29 57.70	 &  14.62   &  14.70   &   13.19  &	      0.053 &	     6900 &	 $-$1.86 &   ....   &	...  &  PS00 \\ 
    CS22950$-$173 &  20 35 31.26, $-$15 53 30.60	 &  14.04   &  14.10   &   12.66  &	      0.039 &	     6800 &	 $-$2.50 &   ....   &	...  &  PS00 \\ 
    CS22960$-$058 &  22 16 05.87, $-$42 26 42.80	 &  13.50   &  13.50   &    ....  &	      0.012 &	     6900 &	 $-$1.99 &   ....   &	...  &  PS00 \\ 
    CS22964$-$074 &  19 49 29.42, $-$39 42 39.40	 &  14.46   &  14.50   &   13.12  &	      0.055 &	     6950 &	 $-$2.30 &   ....   &	...  &  PS00 \\ 
\hline
\hline
\end{tabular}}
 \tablefoot{References: R99: \citet{Ryan1999},   PS00: \citet{ps00}, 	R01:, \citet{Ryan2001},					C05: \citet{Carney2005},
A06: \citet{Asplund2006}, 
M10: \citet{Melendez2010},			
S10: \citet{Sbordone2010}. }
\end{table*}

The sample listed in { Table~\ref{tab:photometry1} and \ref{tab:photometry2} }is composed of stars from large samples presenting various levels of
Li \citep{Ryan1999,ps00,Ryan2001,Carney2005,Asplund2006,Bonifacio2007,Melendez2010,Sbordone2010}. We intentionally choose studies that have an overlap { in order to gain a better handle on the systematic offsets among the different methods they employ. We also} include some of the
well-studied Gaia-ESO Survey benchmark stars \citep{2015A&A...582A..49H, Jofre2014}, and explore how the
temperature scales offset the Li abundances. An example of this is
shown in Fig.~\ref{Lit_spread} where two stars (\object{LP 815$-$43} and
\object{HD 84937}) show temperatures that span $\sim 250$\,K while their [Fe/H] is better constrained (within
$\pm 0.2$\,dex). For A(Li) in HD84937 \citep[][T=6408K]{Melendez2010} a value of 2.32 is derived,
while the study by \citet{Ryan1999} obtained a lower temperature (T$=6160$\,K) and in turn a
lower A(Li)$=2.17$. This means that 250\,K is enough to shift the lithium abundance
above or below the Spite plateau \citep{Spite1982}, but it stays within the uncertainty of Li abundances on the plateau \citep[$\sim 0.09$\,dex according to ][]{Bonifacio2007}.

\subsection{Li and $V-Ks$ as binary/BMP segregators}\label{Licomments}
Since our compilation of Li abundances relies on literature, we note that the different methods
employed may shift the results by up to 250\,K or $\sim 0.2$\,dex in A(Li). { This abundance offset arises because the derived Li value heavily depends on the effective temperature; a parameter that differs when derived from phometry (IRFM) to spectroscopy (e.g., H$\alpha$-fitting), and both of these methods have been used in the studies we compare to.  }
Hence, to
lower this impact of temperature on the absolute value of A(Li) we turn
towards photometry and study the direct impact of the star's blue colour, and later connect this to the lithium abundance { (through stellar spectroscopy)}. 
{ Based on the average uncertainties in photometry ($\pm 0.01$\,mag) and dereddening (0.01 -- 0.1\,mag) taken from the same literature as the BMP sample \citep[e.g.,][]{Melendez2010,Ryan2001}. A typical uncertainty of $0.01$\,mag translates into $\sim 50$\,K.}
We probe the trends using $U-B, B-V, V-Ks,$ and the dereddened values $U-B_0, B-V_0, V-Ks_0$. These colours can be calculated for the majority of BMP stars studied here. The cleanest trends and best way of categorising the BMP stars into groups is obtained using $B-V_0$ and $V-Ks_0$.
\begin{figure}
\includegraphics[scale=0.5]{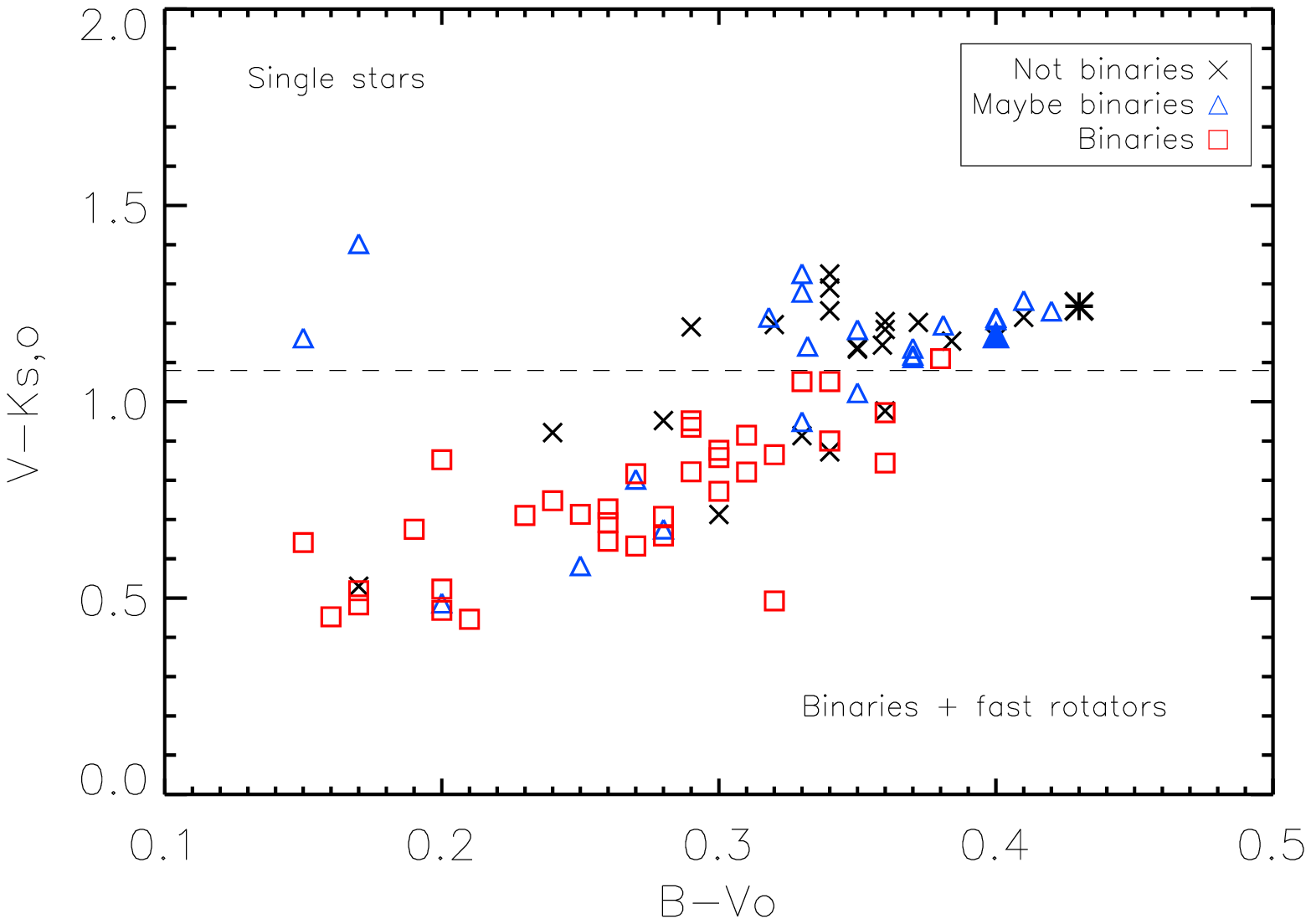}
\includegraphics[scale=0.5]{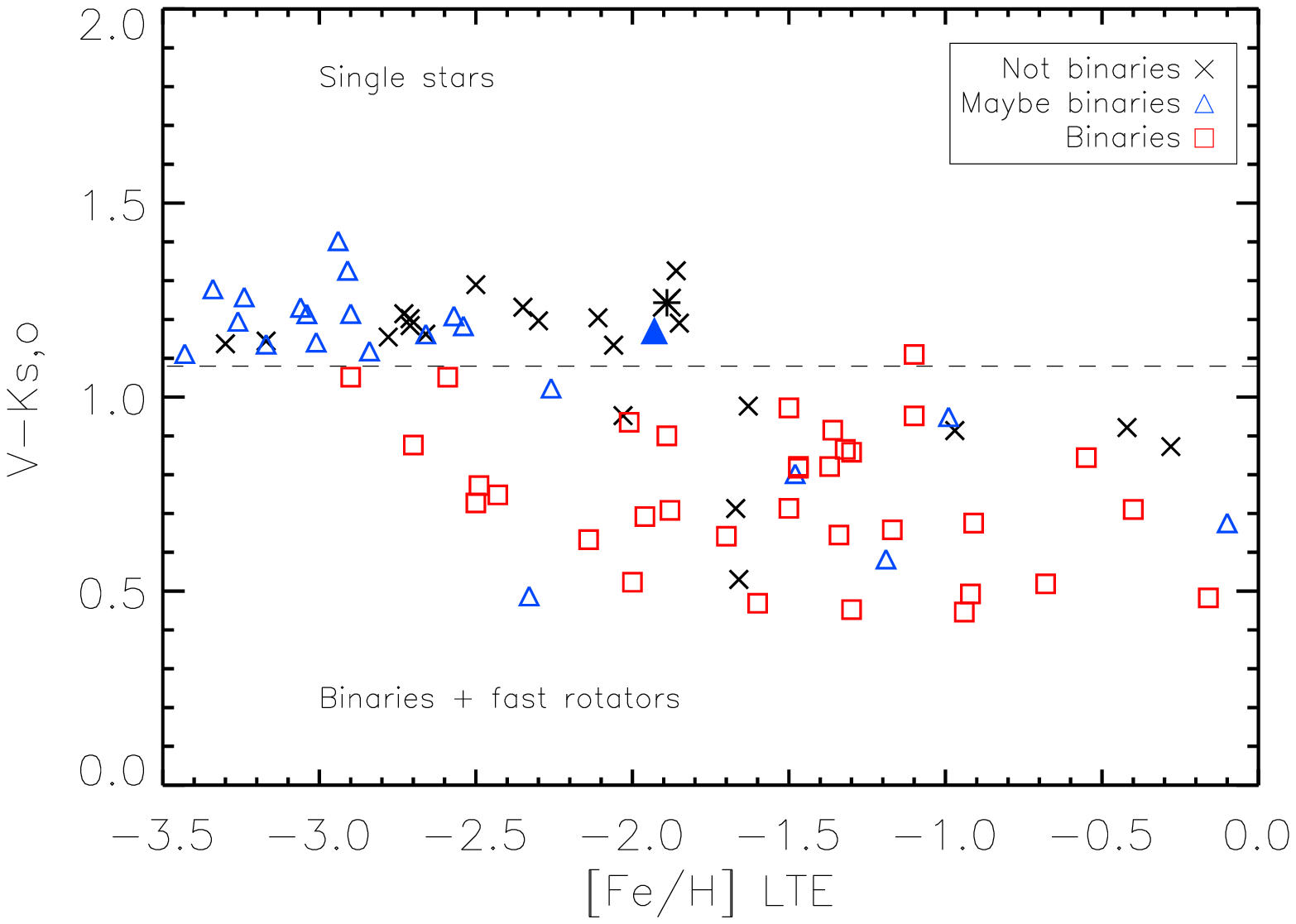}
\centering
\caption{Dereddened $B-V$ vs $V-K_s$ (top) and $V-K_s$ vs [Fe/H] (bottom) for binaries (red squares), maybe binaries (blue triangles) and single stars
(black X). A separation (dashed line) between single stars and binaries is show for $V-Ks_0=1.08$. \cs\ is shown as an asterisc and \cd\ is indicated by a larger, filled triangle.}
\label{BV_VK_Fe}
\end{figure}
The sample of { 80} stars distributed amongst binaries (34 stars, red squares), single stars (23, black `x'), and possible 
binaries (tagged `maybe' owing to a lack of repeat 
radial velocity measurements amounts to 23 stars, blue triangles). The V magnitudes are taken from the respective studies
while the 2MASS K$_s$ magnitudes are based on the SIMBAD references. The dereddening, $E(B-V)$, is based on literature when
provided, otherwise it is the mean S\& F \citep{2011ApJ...737..103S} from the IRSA webpage\footnote{http://irsa.ipac.caltech.edu/applications/DUST/}. Their colour-metallicity distribution is shown in
Fig.~\ref{BV_VK_Fe}. This figure surprisingly shows that all but one binary star fall below $V-Ks_0 = 1.08$. This
corresponds to 97\% of the binary stars falling below this cut with a pollution of 1/3 of the single stars. A similar, but
less clean finding is made for $B-V_0$. { We note that a linear relationship between the sum of the three Ca line equivalent widths and the metallicity in a globular cluster was discovered by \citet{Armandroff1988} using integrated light spectroscopy. With this in mind, we see that} simple metallicity indicators such as the near-IR Ca triplet (CaT) combined with photometry can
be used as a first approach to sort blue stars into binaries and single stars as well as classifying possible
binary candidates into one of these two groups (without needing follow-up observations).  

\begin{figure}
\includegraphics[scale=0.5]{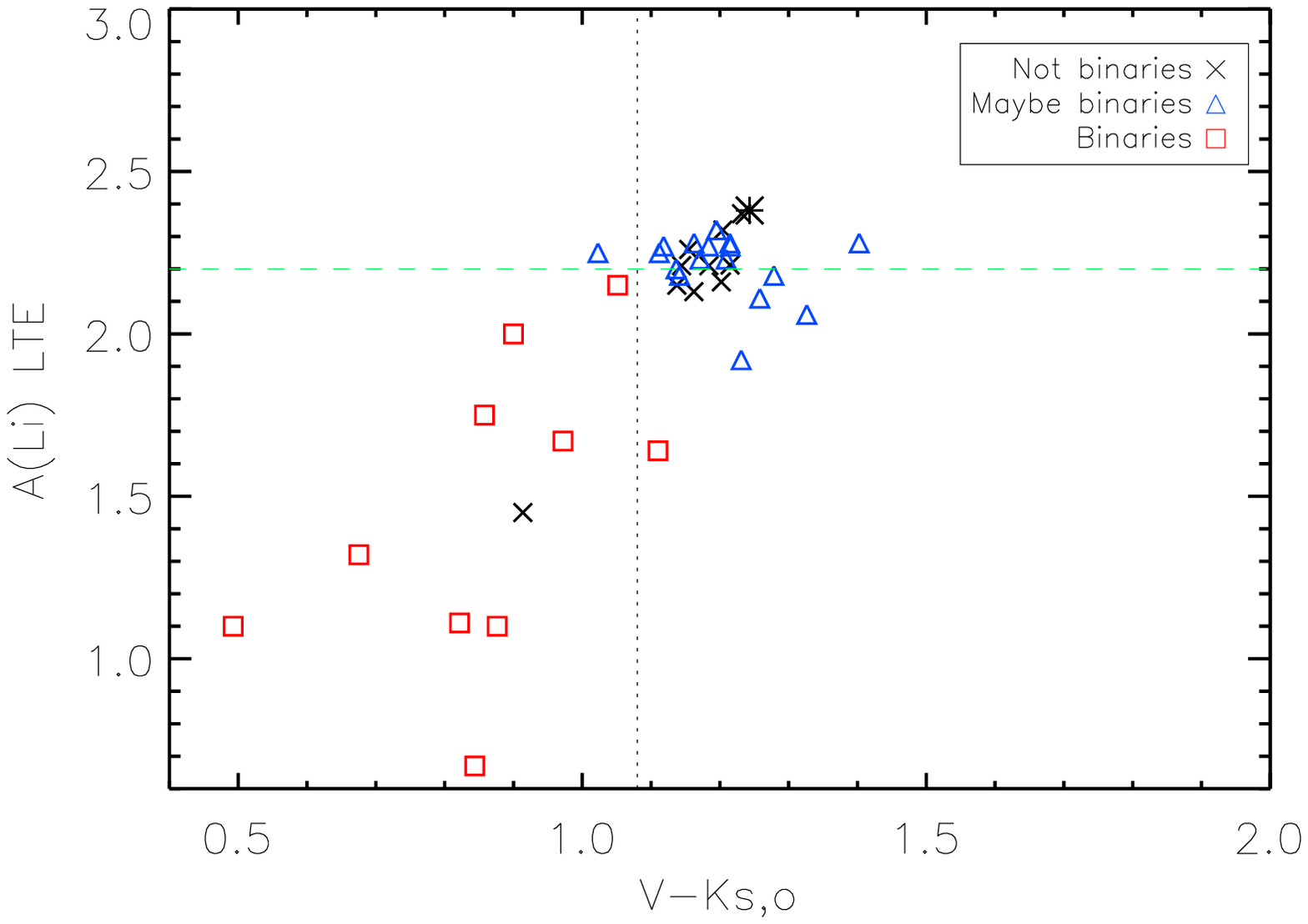}
\includegraphics[scale=0.5]{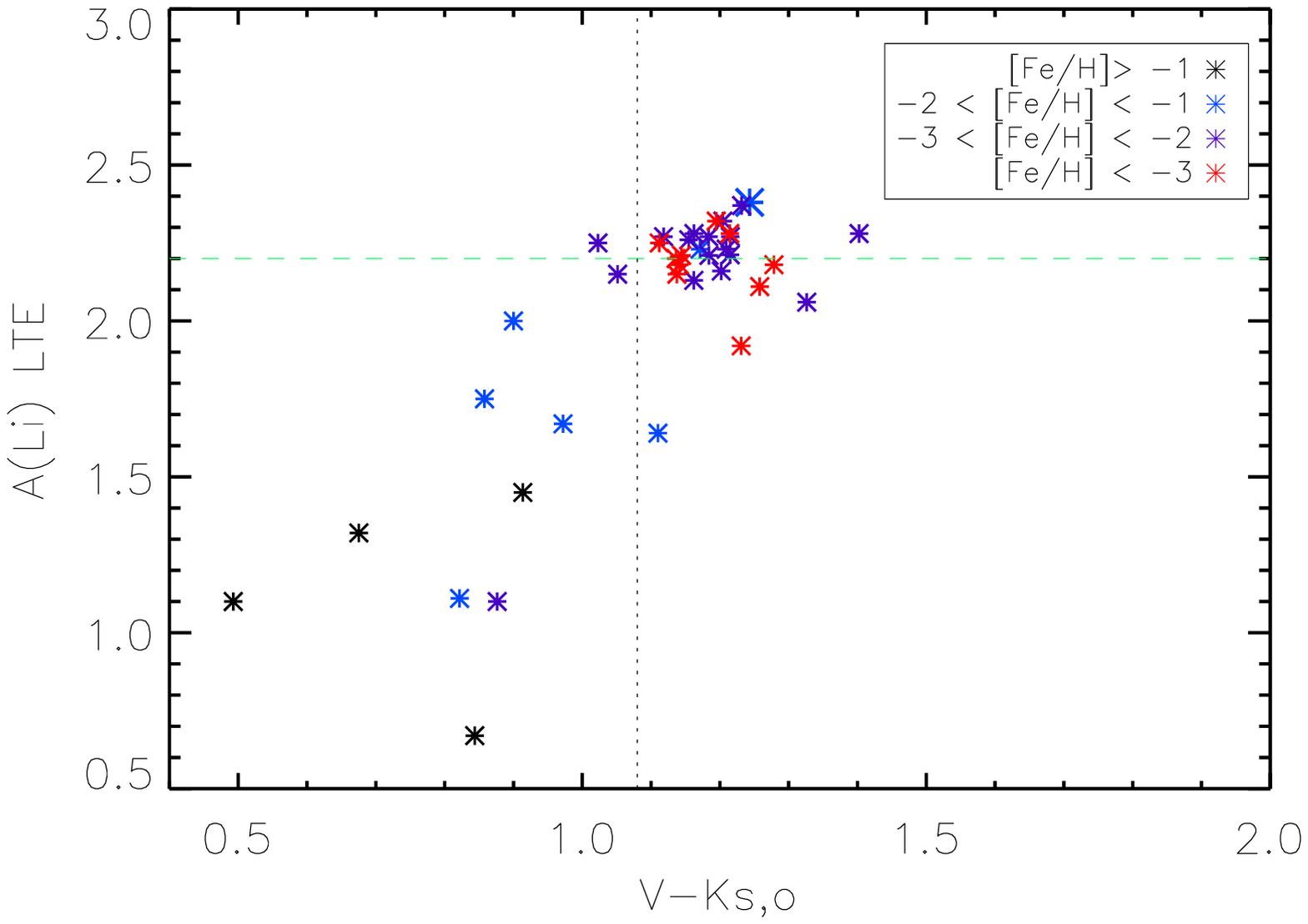}
\centering
\caption{A(Li) vs dereddened $V-Ks$ (legend as in Fig.~\ref{BV_VK_Fe}. Binaries are separated from single stars by
$V-Ks_0$=1.08 and the Spite plateau. The same sample is colour coded by metallicity (bottom panel). Values for \cd\ and \cs\ are from this study with \cs\ shown as a larger asterisc.}
\label{VK0_Li}
\end{figure}
However, these trends become stronger and cleaner once A(Li) is considered.
Figure~\ref{VK0_Li} shows A(Li) as a function of $V-Ks_0$ with a very clean separation of binaries below 1.08 and single stars above this colour cut. From the same figure binaries are seen to fall below the Spite plateau (green long dashed line) while the most metal-poor, single stars are seen to cluster around the Spite plateau, and mainly reside above $V-Ks_0 = 1.08$ (see top panel in
Fig.~\ref{VK0_Li}). The trend is clean and the contamination is low (1 out of 39 stars does not follow the trend). 
This confirms that Li is a useful measurement to separate BMP stars into blue stragglers and metal-poor, single stars \citep{Carney2005}. This method is a very
promising tool 
using $V-Ks_0 = 1.08$ and the Spite plateau as segregators of binary stars. { A slight increase in [Fe/H] with decreasing $V-Ks_0$ is seen in the bottom panel of Fig.~\ref{VK0_Li}.}

\begin{figure}
\includegraphics[scale=0.5]{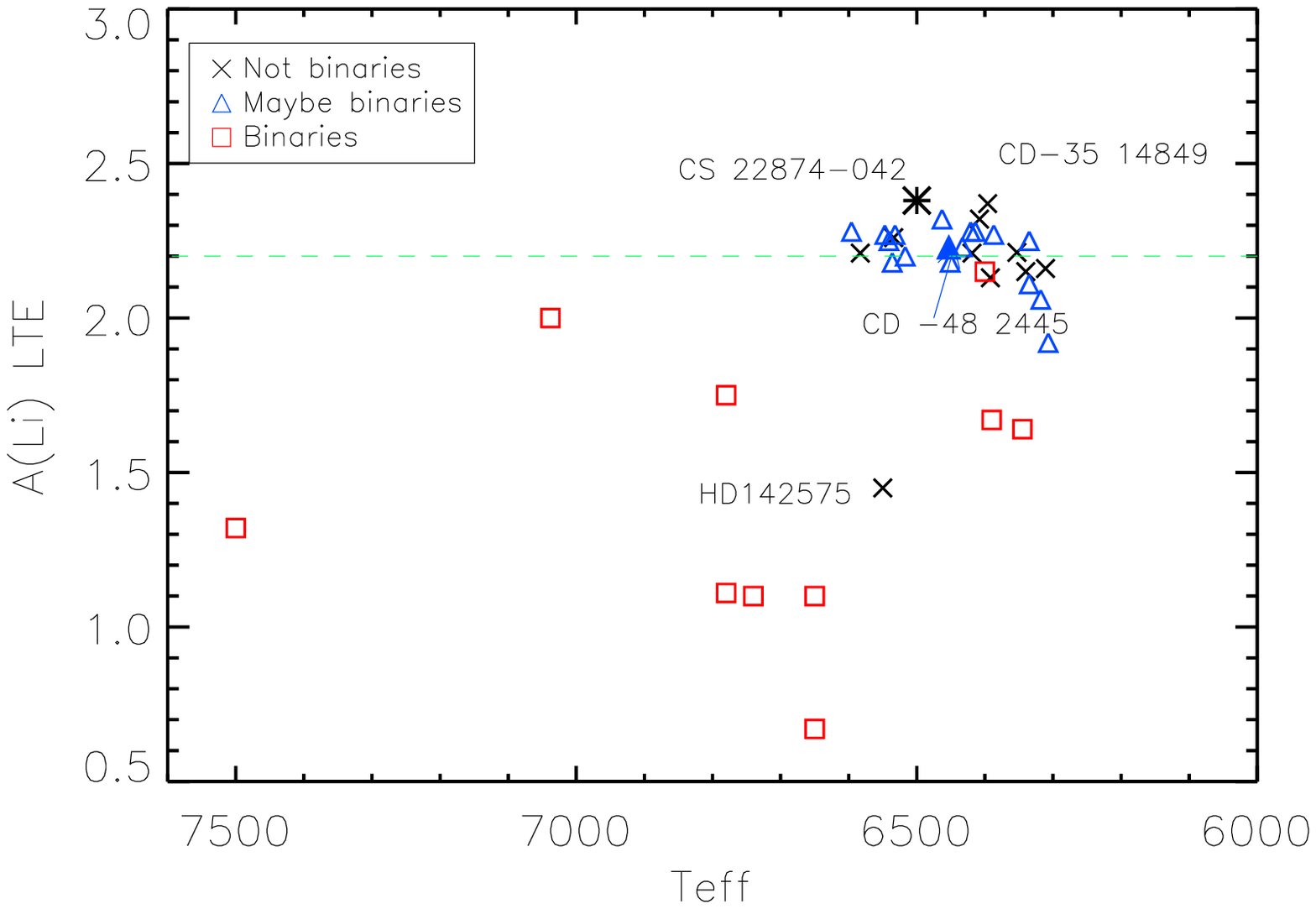}
\includegraphics[scale=0.5]{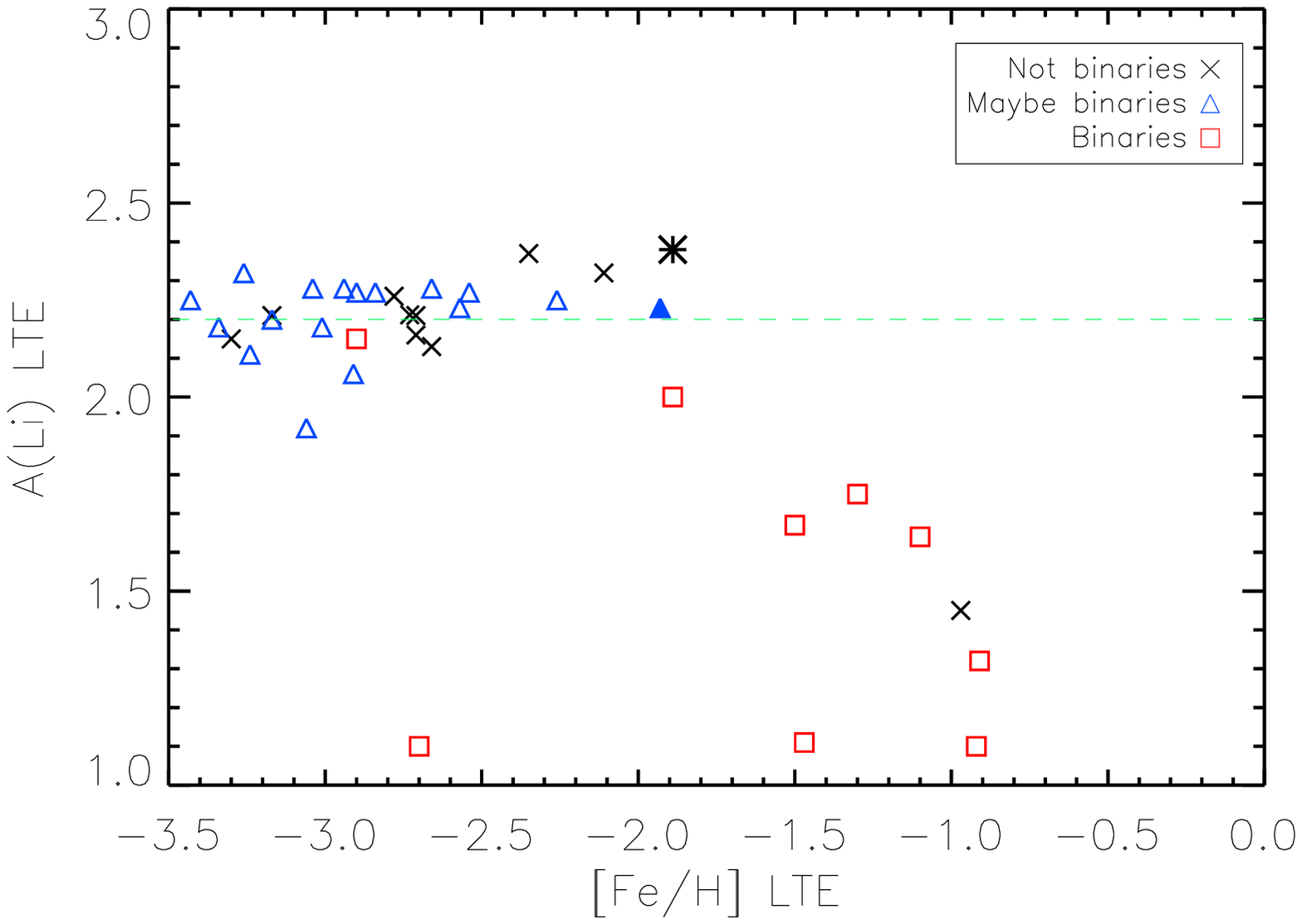}
\centering
\caption{A(Li) vs T and [Fe/H] (legend as in Fig.~\ref{BV_VK_Fe}. Except from one case, the binaries do not reach the Spite
plateau.}
\label{Li_T_Fe}
\end{figure}

Now considering A(Li) as a function of temperature we find outliers above and below the Spite plateau. The Li melt-down from \citet{Sbordone2010} using IRFM in concordance with the methods employed by most studies shown here seem to need a { strong} downward shift in both Li and [Fe/H] to match { these BMP stars}. 
These are less prominent when A(Li) is plotted vs [Fe/H] (see Fig.~\ref{Li_T_Fe}).
The three stars with the highest A(Li) in Table~\ref{tab:photometry1} and \ref{tab:photometry2} { according to \citet{Melendez2010}} are: \object{CD $-$35 14849}, \object{CD $-$48 2445}, and \object{CS 22874$-$042} (shown by a larger asterisc). { However, \citet{Asplund2006} report a lower Li abundance (2.22\,dex) for \object{CD $-$48 2445} and several of the other stars that overlap between the two studies. Our analysis support the lower values and we therefore only} conduct a spectrum analysis the latter two of the three stars, a possible binary and a single star, to investigate their formation mechanism.
Another outlier is the single star HD142575, which \citet{Carney2005} explained by a high rotational velocity decreasing the Li
abundance so that it in our classification could be mistaken as a binary star. This makes blue fast rotating stars possible contaminators in our colour-separation method. Thus, in the following we look into the impact of rotation and
the existence of the so-called Li gap or dip \citep[see, e.g.,][and Sect.~\ref{Lidip}]{Boesgaard1986}.

{ Using the colour cut and Li we thus select \cd\ and \object{HD 142575} as comparison stars. In our high SNR spectroscopic analysis we use their abundances to explore differences and similarities in their formation compared to that of \cs.}

\section{Spectrum analysis of extreme BMP and benchmark stars}\label{data}
{ Here we define a high Li abundance value (Li-rich) as A(Li)$>2.3$\,dex (Spite Plateau plus typical uncertainty $\sim0.09$\,dex) in very metal-poor stars with T$>6300$\,K. Meanwhile, we will refer to Li-normal as 2.2 and Li-low as $<2.1$\,dex.} 
To place  the analysis of \cs\ in context, we will in the following also consider two other BMP stars for comparison:  
HD~142575 (Li poor), and \cd\ (Li rich, see Fig.~\ref{Li_T_Fe} { according to \citet{Melendez2010}}).  The photometric properties are summarised in \tab{bmp_obs}.  A short recap of stars we compare to in our spectroscopic study is presented in Table~\ref{Tab:compsample}. These stars are the outliers or most extreme BMP stars tagged in Fig.~\ref{Li_T_Fe}.
 \begin{table*}
 \caption{Photometric data of the three stars employed in our analysis as taken from Simbad.  
 $B$ and $V$ magnitudes are on the Johnson Cousin system.  For \cs\ they were taken from \citet{2007ApJS..168..277B} while for \cd\ and HD~142575  the values were taken from Simbad. All  $JHK_s$ magnitudes are from the 2MASS catalog, and reddening mean values from  \citet[IRSA, ][]{2011ApJ...737..103S}.}
 \label{bmp_obs}
\hspace{-0.1cm}
\centering
\begin{tabular}{c  c c  c   c c c c  c }
\hline
\hline
\small
Star & RA (J2000.0) & DEC  (J2000.0) & $B$ & $V$ & $J$ & $H$ & $K_s$  &  E$(B-V)$  \\
\hline
\object{CS 22874$-$024}  & 14 38 01.7 & $-$24 58 47 & 14.34& 13.91 & 12.77 & 12.48 & 12.42 &0.076 \\
\object{CD $-$48 02445} & 06 41 26.7 &  $-$48 13 15 &  10.94 & 10.54 & \phantom{1}9.59 &  \phantom{1}9.34 &  \phantom{1}9.29 & 0.071 \\
\object{HD 142575} &  	15 55 02.8 & +05 04 12 &   \phantom{1}9.00 &  \phantom{1}8.62 &  \phantom{1}7.72 &  \phantom{1}7.56 &  \phantom{1}7.51 &  0.056   \\
\hline
\hline
\end{tabular}
\end{table*}
\subsection{\cs}
The spectrum from PS00 is plotted in \fig{li_ps00} and corresponds to 22 stacked, individual exposures. It
clearly shows a strong Li feature highlighting our interest in follow-up observations of this Li-rich BMP star { (according to our convention see Sect.\ref{data})}. 
A high-resolution spectrum of CS22874-042 was obtained on
19/20, August, 2014, using the MIKE echelle spectrograph on the
6.5m Clay telescope at Las Campanas Observatory.
We employed a 0.5x5.0" slit and 2x1 binning of the CCD pixels.
Three exposures were obtained, in order to better eliminate cosmic-ray
events, for a total integration time of 3000\,sec.  During the first two exposures
the guide camera seeing was 0.53", fwhm, but for the third 
integration the seeing worsened somewhat, ranging from 0.67 to 0.89"
fwhm.

The MIKE CCD data were reduced using the pipeline developed by \citet{2003PASP..115..688K}.
The final blue side spectrum provided continuous wavelength coverage from
3360\,\AA~to 5000\,\AA\ , at a resolving power of R~$\sim$57,000.  
At the peak of the orders in the blue-side spectrum, a S/N of 66 per pixel was obtained near 5000\,\AA , increasing to maximum of 98 per pixel at
4679\,\AA , and then steadily declining: at 4000\,\AA\ S/N=64, while at 3364\,\AA\
S/N=9 per pixel.

The red side
spectrum provided continuous coverage from 4950\,\AA~to 9400\,\AA , with 
a resolving power of R$\sim$48,000.  The S/N peaked near 100 per pixel 
at 8590\,\AA , slowly declining with decreasing wavelength: a S/N=87 per pixel 
was obtained at the peak of the H$_{\alpha}$ order, 76 per pixel in the
Na~D order, and 52 per pixel in the Mg~b order.

The RV of \cs\ was re-determined by cross-correlating the spectrum with a template of HD~84937 using the code {\tt iSpec} \citep{2014A&A...569A.111B}. We used this star as template because its stellar parameters  \citep{Jofre2014,  2015A&A...582A..49H} are similar to this BMP star. Moreover, we have a  high-resolution and high-SNR spectrum covering the entire optical range \citep[see][for details]{2014A&A...566A..98B}. The RV obtained from this was  178.5$\pm$0.6 km\,s$^{-1}$ and is fully consistent with the values reported by PS00 { $>14$} years ago.  This consolidates the fact that this star has no significant RV variations, confirming its purported origin as a metal-poor, single, blue star. Owing to its $\alpha-$enhancement ([$\alpha$/Fe] $= 0.35$) it seems unlikely that this star should be an accreted star. We look further into the formation of this star in \sect{spec} and \sect{discussion}.

\begin{table*}
\caption{Stellar parameters of \cs\ derived in different ways. The photometric temperatures were derived using \citet{2010A&A...512A..54C} - C10, \citet{2005ApJ...626..446R} - RM05, and \citep{1996A&A...313..873A} - A96. Method $b$ is iterated to a stronger/smaller uncertainty than $a$. }
 \label{stellarpar}
 \centering
\begin{tabular}{l c c c}
\hline
\hline
Parameter & Values & Tag/Comment & Weight \\
\hline
T$_{ex}$ & 6500/{ 6550}/6850 & a/b/PS00  &  1/0/0  \\
\hline
T$_{H\delta}$ & 6600 &  - &  1  \\
T$_{H\gamma}$ & 6450 & -  &   1 \\
T$_{H\beta}$ & 6350 & log$g$ sensitive  &  0  \\
T$_{H\alpha}$ & 6250 & log$g$ sensitive   &  0  \\
\hline
T$_{B-V}$ & 6700/6450/6600 &  C10/A96/RM05 &  1/0/0  \\
T$_{V-K_s}$ & 6300 &  A96 &   1 \\
T$_{J-K_s}$ & 6450 &  C10 &   1 \\
\hline
T$_{final}$ & $6500 \pm 100$ &   &    \\
\hline
log$g_{Fe}$  & 4.5  & & 0.8   \\
log$g_{Mg}$  & 4.6  & & 0.2     \\
\hline
log$g_{final}$  & $4.5 \pm 0.1$ &  &   \\
\hline
$[$FeI$/$H$]$  & { $-1.95$ }&  & 0.5 \\
$[$FeII$/$H$]$ & $-1.91$ &  & 0.5 \\
\hline
$[$Fe$/$H$]_{final}$ & { $-1.9 \pm 0.1$} & & \\
\hline
V$_{mic}$  & { $1.9 \pm 0.1$} km/s & Fe I lines & \\
Vsin$i$  & { $2.7\pm0.3$} km/s  & Profile matching & \\
RV$_{helio}$ & { $178.5\pm0.6$} km/s & Cross correlation  & \\
\hline
\hline
 \end{tabular}
 \end{table*}

 \subsection{\cd}\label{bdstar}
 
 This star has been included as a comparative star as its stellar parameters are very similar to the parameters we derived for \cs\ (see \sect{params}). We adopted the stellar parameters from \cite{Melendez2010} who reported  \teff=6453\,K, \logg=4.25, \feh$=-1.93$, \vmic=$1.5~\mathrm{km\,s^{-1}}$  and A(Li) = 2.38 when assuming local thermodynamical equilibrium (LTE).  In the following we compute all stellar abundances using these parameters. { We note that \citet{Asplund2006} found lower values for the temperature and Li in this star. We therefore check if the absoprtion lines (including the Balmer lines) are well fit with the higher adopted temperature in order to reassess the Li value (see Table~\ref{Tab:compsample}, Fig.~\ref{profiles}, and \sect{abund} for further details on our derived abundances.}
 
 The spectrum of this star has been obtained with the UVES instrument in the BLUE346 and REDL580  settings in November 2012\footnote{Programme ID: 090.B-0605} and thus its reduced spectrum is publicly available. The SNR is 60 
 per pixel in the blue and 100 in the red part, and the resolution is of 60,000. We corrected its RV using {\tt iSpec} in the same way as for \cs\ obtaining a heliocentric velocity of $305 \pm 0.23~\mathrm{km\,s^{-1}}$ for \cd. 
 
 The spectrum analysed in  \cite{Asplund2006} was taken in 2002 and the authors derived an RV of { $319.2\pm0.3~\mathrm{km\,s^{-1}}$ and noted that this star showed no evidence of having a companion}, while \cite{Ryan1991} measured an RV of {  $338\pm7~\mathrm{km\,s^{-1}}$} and \cite{1979AJ.....84.1553A} reported an RV of $301\pm1.4~\mathrm{km\,s^{-1}}$ determined from a spectrum taken in 1976.  { Taking the combined uncertainties into account ($<10~\mathrm{km\,s^{-1}}$, \cd\ seems be a binary star given the difference of $37~\mathrm{km\,s^{-1}}$ over the 40 years timespan between these different works. Our colour cut ($V-Ks_0=1.08$) confirms} the suggested binary nature of \cd. Hence, we look into mass transfer signatures in \sect{sect:AGB} { as this star is an excellent candidate of a binary system now containing a low-mass ($<0.55$M$_{\odot}$) white dwarf}.
  
\subsection{HD~142575}
This star has been included in this analysis because it is part of the RV constant ($-${ 65}km\,s$^{-1}$)  BMP stars from \citet{Carney2005}, with measured Li abundance of A(Li) = 1.45; it has a temperature of 6700\,K placing it within the Li-dip. This BMP star has a rotational velocity of $\ v sin i = 13.0~\mathrm{km\,s^{-1}}$ and a metallicity of $-0.97$\,dex \citep{Carney2005}. 
{ The spectrum of this star was} obtained with the UVES instrument with the in BLUE346, REDL580 and REDL860 gratings in 2006-06-11\footnote{Programme ID: 077.B-0507}.

  \begin{table*}
 \caption{Stellar parameters  and Li-abundance of the stars we compare to with values taken from literature. An `*' indicates 3D analysis. }
 \label{Tab:compsample}
 \centering
\begin{tabular}{l c c c c c c c}
\hline
\hline
star	&	Teff	&	logg &	[Fe/H] & vmic & vsini & A(Li) & Reference\\
 & (K) & (dex) & (dex) & (km/s) & (km/s) & \\
\hline
\object{BPS CS22874$-$042} & 6500 & 4.5 & { $-1.9$} & { 1.9} & 2.7 & 2.38 & This work \\
\hline
\object{BPS CS22874$-$042} & 6900 & 4.5 & $-1.53$ & 2.0 & 8.0 & -- &PS00\\
\object{CD $-$48~2445} & 6222/6453 & 4.25/4.25 & $-1.93/-1.93$& 1.5/1.5 & 2.3*/-- & 2.22/2.38 & A06/M10 \\
\object{HD 142575} & 6550/6700 & --/3.6 & $-0.97/-0.9$ & --/1.7 &  13.0 & 1.45 & C05/F00 \\
\object{HD 106038} & 5905/5950 & 4.3/4.3 & $-1.35/-1.44$ & 1.2/1.1 & 0.5/-& 2.48/-- & A06/H12 \\  
\object{HD 84937} & 6356/6300 & 4.06/-- & $-2.03/-2.18$ & 1.3/-& 5.2/4.8& 2.23 &J15/C05\\
\object{BPS CS 22950$-$173} & 6335/6506 & 4.2/4.5 & $-2.78/-2.61$ & 1.4/1.4 & -- &  2.20&S10 (min/max) \\  
\object{LP 815$-$43} & 6453/6630 & 3.8/4.1 & $-2.88/-2.77$ & 1.7/1.7 & -- / --  & 2.23/2.16 & S10/A06 \\
\hline
\hline
 \end{tabular}
 \tablefoot{References:  F00: \citet[][]{Fulbright2000}, PS00: \citet[][]{ps00}, 										
C05: \citet[][]{Carney2005},
A06: \citet{Asplund2006}, 
M10: \citet{Melendez2010},			
S08: \citet{Smiljanic2008},
S10: \citet{Sbordone2010}, 											
H12: \citet{Hansen2012},
J15: \cite{2015A&A...582A..81J}. }
 \end{table*} 

\section{Stellar parameters and abundances}\label{spec}
In this section we present the new results from our stellar parameter determination and the spectral analysis of the two BMP stars under study. 

\subsection{New atmospheric parameters of \cs} \label{params}
From our new spectra we re-determined the stellar parameters and, while we found  different parameters from PS00, we  still support the view that this \cs\  is a BMP star. Since the final Li abundance is very sensitive to the effective temperature, we determined the stellar parameters using various methods. We considered photometry for the T$_{\rm eff}$,  ionisation and excitation equilibrium from the EWs  of iron lines, as well as synthesis of several spectral features such as  the Mg triplet and the Balmer wings,  and various iron line profiles. A summary of our results can be found in \tab{stellarpar}.  Not all temperature or gravity indicators are equally good and we therefore assign weights to the methods we employ. Here we use values between 1 (full weight) and 0 (method discarded) to indicate the fraction with which the value enters the final averaged value. Below we describe  each of these analyses in greater detail. \\ %

\noindent {\it Photometric effective temperature: } We used three different Infrared Flux Method (IRFM) calibrations to determine the temperature, namely \citet{2010A&A...512A..54C}, \citet{2005ApJ...626..446R}, and \citep{1996A&A...313..873A} for dwarfs and subgiants. These methods have also been used in the literature samples we compared to. We calculated the effective temperature using different colours ($B-V$, $V-K$, $J-K$), employing the three calibrations mentioned above, where applicable.   The reddening $\mathrm{E}(B-V)$ was found in the IRSA Dust maps of \citet{2011ApJ...737..103S} and \citet{1998ApJ...500..525S}, and we calculated the temperature using both (mean) values (0.075\,$\mathrm{mag}$ for \cs\ and 0.0878\,$\mathrm{mag}$ for \cd ). 
This lead to an $\mathrm{E}(B-V)$ difference of $\sim 0.01~\mathrm{mag}$ which translates into a temperature uncertainty of $\sim50~\mathrm{K}$. 
We therefore round off all photometric temperatures to the nearest 50K. The values listed in Table~\ref{stellarpar} are based on the $\mathrm{E}(B-V)$ from \citet{2011ApJ...737..103S} and the $B,V$ photometry from \citet{2007ApJS..168..277B} and 2MASS from \citet{2006AJ....131.1163S}. \\

\noindent{\it Effective temperature from wings of Balmer lines:} With {\tt iSpec} we determined \teff\ using the wings of the Balmer lines: $\mathrm{H}_\alpha$, $\mathrm{H}_\beta$, $\mathrm{H}_\gamma$ and $\mathrm{H}_\delta$. The results can be seen in \tab{stellarpar}. As expected, NLTE effects are important for metal-poor dwarfs, and depending on the stellar parameters the LTE temperature from, e.g., H$\alpha$ may be a few percent too low  \citep[typically 50-100\,K;][]{2007A&A...466..327B}. \\

\noindent {\it Surface gravity from Mg triplet}: The surface gravity of our target was confirmed with {\tt iSpec} by fitting the Mg triplet lines. For this determination we tested different values of effective temperature and metallicity around our accepted 
value of 6500\,K and $-$1.8\,dex, respectively (see below). In all cases the gravities yielded between 4.3 and 4.6, supporting the gravity of 4.5 reported in PS00.  This rules out the possibility of our star being a BHB or an RR Lyrae star, which have lower gravities. \\

\begin{figure*}
\includegraphics[width=1\hsize]{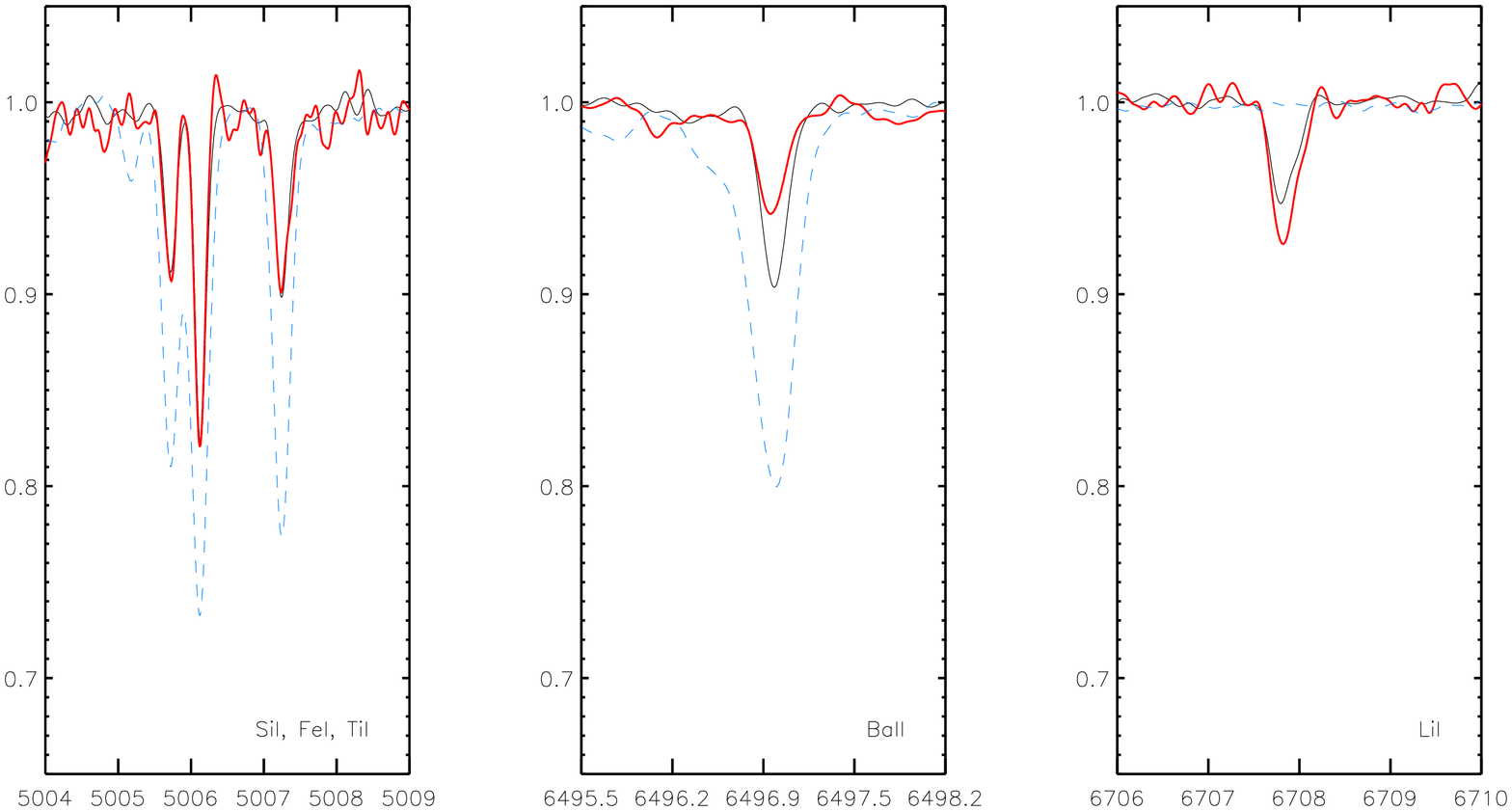}
\includegraphics[width=1\hsize]{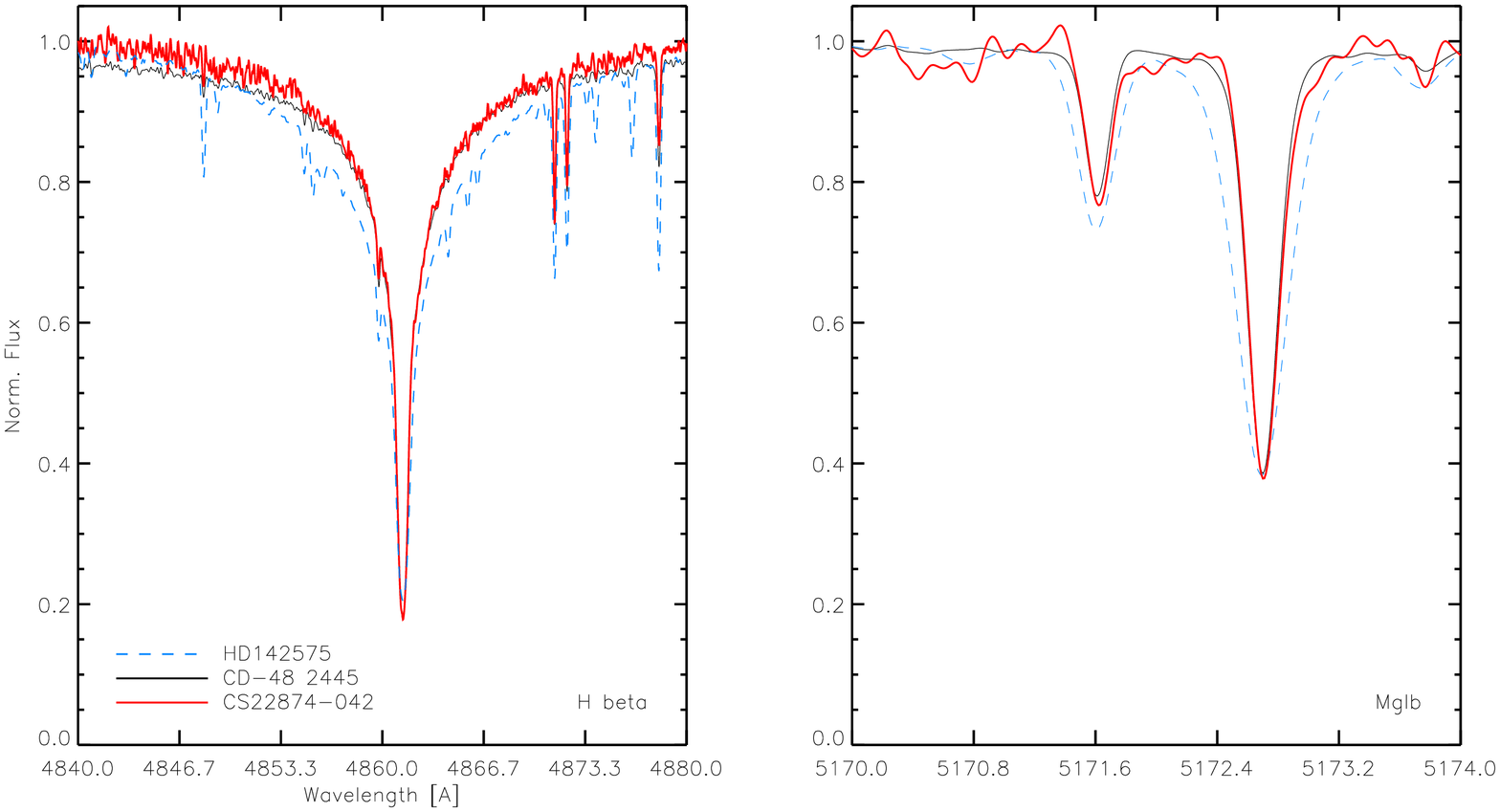}
\centering
\caption{Illustrative line profiles in different regions of the spectra. 
The stellar parameters of \cd\ and \cs, as well as $\alpha-$abundances are comparable, while the abundance of Ba and Li are different. 
HD~142575 has a higher metallicity, temperature and rotational velocity with respect to the other BMP stars. Li in this star can not be measured here. }
\label{profiles}
\end{figure*}

\noindent {\it Parameters from iron line syntheses:} We used the {\tt iSpec} code to fit 150 Fe lines (amongst them, 17 lines are of the ionised species) ranging from 3440 to 7900 \AA.  For the regions outside the line list used within the Gaia-ESO Survey \citep{2015PhyS...90e4010H}, we considered the atomic information provided by VALD \citep{2011BaltA..20..503K}, as provided in the wrapper of {\tt iSpec}. { This method is based on a large number of lines and we therefore assign the resulting temperature the largest weight.}
By fixing $\log g = 4.5$  we obtained a temperature of 6500~K and a metallicity of $-1.8$ dex.  When letting $\log g$ vary we obtained  a slightly lower temperature of $\sim$6450~K, 
but recovered the same  metallicity and a gravity of 4.2. \\

\noindent {\it Parameters from equivalent widths  of iron lines:} {For this test we used Fe lines below 5000\AA\, from \citet{Hansen2012}. In total { 35 Fe~I lines ($\sigma_{\mathrm{[Fe~I/H]}}=0.1$\,dex) and 15 Fe~II ($\sigma_{\mathrm{[Fe~II/H]}}=0.06$\,dex)} lines were used to determine the stellar parameters.
Of these,  19 Fe~I lines and 8 Fe~II lines overlap with the lines in PS00. For most of the Fe~I lines the atomic data agrees to the third digit with their line list values, but for the Fe~II lines there is a (slight) difference in $\log gf$ 
values. This difference can be up to 0.05 but is in most cases around 0.01 or below. The EWs were measured in normalised spectra using IRAF, and the stellar parameters were determined by manually interpolating them in MOOG \citep{1973PhDT.......180S}, the same spectrum synthesis code as used in PS00, but here adopting a more recent version (v. 2014).  

The temperature was set by requiring that all Fe lines 
give the same Fe abundance regardless of excitation potential. The interpolation was stopped when the slope of the trend was less than $\pm 0.05~\mathrm{dex/K}$. 
A similar approach, enforcing ionisation equilibrium between abundances from Fe~I and Fe~II, was applied to fix the gravity. Once the Fe~I and II abundances agreed within 0.05~dex, the iteration stopped. Since ionisation balance is easily achieved and the lines show a low internal line-to-line scatter ($\sim0.1$\,dex) { in good agreement with the {\tt iSpec} line synthesis method}. The metallicity was adopted as an average of all Fe~I and II abundances (as these agree owing to the forced ionisation and excitation equilibrium). { Finally, the microturbulence was determined  minimizing the equivalent width trend with Fe~II abundances, allowing for a 
variation in slope of up to $\pm 0.05~\mathrm{dex/km\,s}^{-1}$. 
The spectroscopically constrained parameters with this method  are (T$_{\rm eff}$, log\,$g$, [Fe/H], $v_{\rm mic}$): 6550K, 4.5 dex, $-1.9$\,dex, 1.9\,km\,s$^{-1}$ (see Table~\ref{stellarpar})}. 

The constrained set of values agree well with the averaged photometric values. However, this temperature is still lower than what was found in PS00. 
Therefore, we carried out the same analysis using only the Fe lines and atomic data from PS00.
When using their lines (38 Fe~I and 8 Fe~II) we find a $\log g$ value that agrees to within 0.1\,dex, a [Fe/H] within 0.07\,dex, and a T$_{\rm eff}$ within 50\,K compared to their published values. The $v_{\rm mic}$ agrees within 0.1\, km\,s$^{-1}$. The choice of lines, spectrum quality, and continuum placement lead to the slight differences in EWs, and in turn abundances, which can explain the differences we find. We note that the spectroscopic temperature may be biased by the few low excitation Fe~I lines measurable.   \\

\noindent { In summary, we derived the fiducial set of stellar parameters (T$_{\rm eff}$/log\,$g$/[Fe/H]/$v_{\rm mic}$/v\,sin\,$i$): \\
{ $6500\pm100$ K/$4.5\pm0.1$ dex/$-1.9\pm0.1$ dex/$1.9\pm0.1$ km\,s$^{-1}$/$2.7\pm0.3$ km\,s$^{-1}$. 
The stated v\,sin\,$i$ could be a placeholder for an unresolved macroturbulence. With these low velocities and the limitation in spectrum quality, we cannot distinguish between v\,sin\,$i$ and v$_{macro}$.}}
The summary of the different results can be found in \tab{stellarpar}. 

We note that the parameters of \cd\ (see \sect{bdstar}) are very similar to our final ones for \cs\ { ($\sim \pm 50$K$, 0.25, 0.03, 0.4$km\,s$^{-1}$)}. Their similar parameters can be confirmed in \fig{profiles}, where we plot some regions of the spectra of the three BMP stars, sensitive to 
stellar parameters as described in \sect{data}. The wings of the Balmer line indicate that the temperatures of \cs\ and \cd\ are comparable, while the temperature of \hd\ is higher. The wings of the Mg line also confirm that the surface gravities of all stars are those of dwarfs, since lower gravities create narrower Mg wings.  In the top left panel we plot the profiles of three (Fe, Si and Ti) lines .
This shows that  \cs\ and \cd\ not only have comparable stellar parameters, but also $\alpha-$ and iron-peak abundances which allows for a very direct comparison of the derived abundances 
in \cs\ and \cd. The main differences between the chemical composition of these two stars are found for the heaviest (s-process) and lightest (Li) elements under study (see Ba and Li panel of Fig.~\ref{profiles}).
\begin{table}{}
\caption{Line list used in our analysis. Full version available online, a portion is shown here for guidance. } \label{info_lines}
 \centering
\begin{tabular}{c c c c c}
\hline
\hline
Wavelength [\AA]& Species & $\log gf$ &E.P. [eV] & Reference  \\
\hline
7771.940  &   8.0  &    0.370  &    9.140  &  WIE    \\
7774.170  &   8.0  &    0.220  &    9.140  &  WIE    \\
7775.390  &   8.0  &    0.000  &    9.140  &  WIE    \\
4057.505  &  12.0  &   $-$1.201  &    4.343  &  SNE    \\
4167.271  &  12.0  &   $-$1.004  &    4.343  &  SNE    \\
4702.991  &  12.0  &   $-$0.666  &    4.343  &  SNE    \\
5528.405  &  12.0  &   $-$0.620  &    4.343  &  SNE    \\
...       & ... & ... & ... & ...\\
\hline
\hline
 \end{tabular}
\tablefoot{References correspond to WIE: \citet{1996atpc.book.....W}; SNE: \citet{2014ApJS..214...26S}, LAW: \citet{2001ApJ...556..452L,2001ApJ...563.1075L}; SOB: \citet{2007ApJ...667.1267S}; YAN: \citet{1998PhRvA..57.1652Y}.}
 \end{table}

\subsection{Chemical abundance determination}\label{abund}

We derived the abundances of 17 elements that display detectable lines in the covered wavelength range of the spectrum of \cs.  
The abundances were derived using MOOG and the line lists from \citet{2014ApJS..214...26S}; 
the atomic data are given in { the online}
 \tab{info_lines}.  
The results for the final abundances of both stars are listed in Table~\ref{tab:abun}.

Among the 17 element of which we measured abundances,  Mg, Ca, Sc, Ti, Cr, Mn, Sr and Ba have been measured by PS00 for \cs\ as well.  Contrary to PS00, we find an $\alpha-$enhancement of \cs\  normal of metal-poor halo stars ([$\alpha$/Fe]$=0.35\pm 0.09$), while PS00 found that \cs\ was `$\alpha-$poor' ([$\alpha$/Fe] $\sim 0.2$) and thus a good candidate for an (intermediate-age) star with extragalactic origin.  
This is due to the lower temperature and metallicity determined by us with respect to the values of PS00.  Regarding the iron-peak elements Sc, Cr and Mn, we obtain a slight enhancement of [Sc/Fe] = 0.3 dex and [Cr/Fe] $= 0.2$\,dex, while  PS00 obtained a similar enhancement for Sc but an underabundance of Cr of $-$0.15 dex. This can be due to differences in line lists like hyperfine splitting or other updates such as line blends. However, we mainly assign this to the better SNR of our new spectrum around the Cr lines compared  to PS00. Manganese is similarly underabundant in both studies.  Finally, regarding the heavy $s-$process elements Ba and La, we obtain a good agreement of Ba abundances with solar values while La is only an upper limit. A moderate enhancement for Sr is derived, in contrast to PS00 who obtained [Sr/Fe] $= -0.4$. This is in part due to the updated atomic data we used for Sr \citep{Bergemann2012, Hansen2013} and in part owing to the different stellar parameters and spectrum quality especially in the blue wavelength range.

With our higher quality spectrum, we were able to measure elements that were not reported by PS00, namely Li, O, Na, Al, Si, V, Co, Ni, Y, and Zr. Like Mn, Al is underabundant. As for La and Ba, a slight weak $s-$process enhancement is found for Sr, Y, and Zr. Silicon and oxygen behave like the rest of the $\alpha-$elements, that is, with the normal enhancement of halo stars, and Co and Ni have solar values. An interesting case is V, which, at 0.3\,dex,  is rather enhanced for an iron-peak element. The line, however, is weak, noisy, and the abundance should be treated with caution.

Finally, the most surprising abundance value is that of Li with A(Li) =  2.38$\pm$0.10\,dex, which remains  enhanced with respect to Spite plateau value, even after our re-determination of the stellar parameters. 
{  
While recent studies attest to an increasing Li with increasing metallicity,
studies of globular clusters at similar metallicity to this star
find a very homogeneous Li-abundance; e.g., NGC 6397 at [Fe/H]=$-2.0$ dex
shows A(Li)=2.25 (Lind et al. 2009) over a broad range of evolutionary 
stages. 
In the metal-poor regime, Bonifacio et al. (2007) state a 1$\sigma$-scatter 
in A(Li) of 0.09\,dex so that CS22874-042 can in fact be considered 
a 2$\sigma$ outlier (modulo our measurement error)}.

 Furthermore, the star presents no detectable $r-$process element lines in its spectrum. This may however be an observational bias owing to the fact that the star is a hot dwarf preventing us from detecting, e.g., the Eu, Sm, Gd, and Dy lines. For example, the EW of the generally strong Eu line at 4129\,\AA\ lies below 3\,m\AA\ at these stellar parameters for a solar Eu abundance. This is at the noise level of our spectrum and thus, detecting very low abundances of Eu for our BMP star is not possible within our spectrum.

For the comparison star, \cd,  the Li abundances have been reported in the literature, with a plateau value of 2.2 \citep{Asplund2006} to 2.38 \citep{Melendez2010}. In comparison we derive an A(Li) of 2.23 agreeing well within the combined errors with the measurement of \cite{Asplund2006}. Here we report abundances for the same 17 elements that were also measured in \cs.  \cd\ shows a slightly higher  [$\alpha$/Fe] $= 0.39\pm 0.14$, but a similarly low Al and Mn abundance, and a general increased level of the $s-$process elements Sr, Zr, Ba, and La. Interestingly, this hot star has an $r-$process enhancement  of [Eu/Fe] = 0.85.

 \begin{table}
 \caption{Final elemental abundances measured for CS 2287$-$042 and CD$-$48 2445. The uncertainty ($\sigma$) covers to the line-to-line scatter and continuum placement. }
 \label{tab:abun}
 \centering
\begin{tabular}{c c c  |c  c }
\hline 
\hline
& \cs& & \cd& \\
\hline
 Elem. & [X/Fe] & $\sigma$  &  [X/Fe] & $\sigma$  \\
\hline
  Li I  &  2.38   &     0.10   &  	       2.23	 & 	-     \\ 
  CH    &  0.0    &     0.2    & 	       <0.0	 & 	-     \\
  O I   &  0.45   &     0.10   &    	      --	 & 	--      \\   
  Na I  &  0.22   &     0.05    &    	       0.19	 & 	  \\ 
  Mg I  &  0.35   &     0.10   & 	       0.55	 &   0.05   \\     
  Al I  &$-0.85$  &     0.05   &    	      $-0.83$	 &   0.09    \\ 
  Si I  &  0.25   &     0.10   &    	       0.20	 &   0.10        \\ 
  Ca I  &  0.28   &     0.10   & 	       0.38	 &   0.0      \\ 
  Sc II &  0.30   &     0.05   &    	       0.26	 &   0.05     \\ 
  Ti I  &  0.46   &     0.15   & 	       --	 & 	--    \\ 
  Ti II &  0.37   &     0.10   & 	       0.42	 &   0.04     \\ 
  V  II &  0.30   &     0.10   &    	       0.27	 &   0.2       \\     
  Cr I  &  0.20   &     0.05   &   	       $-0.05$	 &   0.05     \\ 
  Mn I  &$-0.45$  &     0.05   &    	       $-0.45$	 &   0.01     \\      
  Co I  &  0.00   &     0.05   &    	        0.11	 &   0.04    \\ 
  Ni I  &$-0.05$  &     0.05   & 	      $ <0.10$	 &    --     \\ 
  Sr II &  0.15   &     0.10   &    	       0.50	 &   0.15       \\ 
  Y II  &  0.15   &     0.15   &    	       0.05	 &   0.05     \\ 
  Zr II &  $<1$   &      --    &    	       0.70	 &   0.15    \\ 
  Ba II &  0.0    &     0.10   &    	       0.24	 &   0.04    \\ 
  La II &  $<0.5$ &      --    &               0.70      &   0.10  \\ 
  Eu II &  --   &      --  &    	         0.85	 &   0.1      \\ 
\hline
\hline
 \end{tabular}
 \end{table}

\subsection{Uncertainties}\label{uncertainties}
To assess the uncertainties due to stellar parameters,  we determined the abundances in the same way as in \citet{2015A&A...582A..81J}, using {\tt iSpec}. To wit, 
we determined the abundances eight times, each one considering a different value of the stellar parameters given their uncertainty. The results for each star are indicated in \tab{tab:abu_fin}. The error due to uncertainty in each of the stellar parameters are given as $\Delta$(Teff), $\Delta (\log g)$, $\Delta$([Fe/H]) and $\Delta$(vmic), respectively. 
 \begin{table*}
 \caption{Final uncertainties in the abundances of CS 22874-042.  The variation obtained considering the uncertainties in stellar parameters are given under the variable $\Delta$ for T, logg, [Fe/H], and $\xi$, respectively. The last column indicates the number of lines used for each element. }
 \label{tab:abu_fin}
 \centering
\begin{tabular}{c  c c c c| c  c c  c c}
\hline
\hline
\multicolumn{5}{c}{CS 22874-042}	& \multicolumn{5}{c}{CD $-48$ 2445}	\\
\hline
Element & $\Delta$(Teff) &   $\Delta$(logg) & $\Delta$([Fe/H]) &  $\Delta (\xi)$ &  $\Delta$(Teff) &   $\Delta$(logg) & $\Delta$([Fe/H]) &  $\Delta (\xi)$ & N lines\\
\hline
Li &  0.131 &  0.000 &  0.010 &  0.015 &     0.07 &   0.00 &   0.00 &   0.00 & 1  \\
 C &   --   &   --   &   --   &   --   &     0.50 &   0.13 &   0.61 &   0.01 & 2  \\
 O &  0.125 &  0.076 &  0.009 &  0.013 &      --  &   --   &    --  &    --  & 0  \\
Na &  0.171 &  0.052 &  0.003 &  0.005 &     0.08 &   0.02 &   0.00 &   0.03 & 2  \\
Mg &  0.073 &  0.003 &  0.004 &  0.005 &     0.04 &   0.00 &   0.00 &   0.00 & 3  \\
Al &  0.123 &  0.000 &  0.025 &  0.043 &     0.06 &   0.00 &   0.01 &   0.02 & 2  \\
Si &   --   &   --   &   --   &   --   &     0.08 &   0.03 &   0.00 &   0.03 & 1  \\
Ca &  0.170 &  0.051 &  0.005 &  0.007 &     0.09 &   0.03 &   0.00 &   0.01 & 4  \\
Sc &  0.094 &  0.071 &  0.004 &  0.006 &     0.05 &   0.03 &   0.00 &   0.01 & 3  \\
Ti &  0.049 &  0.074 &  0.002 &  0.002 &     0.03 &   0.04 &   0.00 &   0.00 & 1  \\
 V &  0.042 &  0.074 &  0.005 &  0.009 &     0.02 &   0.04 &   0.00 &   0.00 & 1  \\
Cr &  0.187 &  0.008 &  0.007 &  0.010 &     0.09 &   0.00 &   0.00 &   0.05 & 2  \\
Mn &  0.186 &  0.002 &  0.007 &  0.010 &     0.10 &   0.00 &   0.00 &   0.01 & 3  \\
Co &  0.143 &  0.010 &  0.010 &  0.015 &     0.07 &   0.00 &   0.00 &   0.00 & 2  \\
Ni &  0.146 &  0.004 &  0.010 &  0.015 &     0.07 &   0.00 &   0.00 &   0.00 & 1  \\
Sr &  0.132 &  0.044 &  0.017 &  0.028 &     0.07 &   0.02 &   0.01 &   0.08 & 2  \\
 Y &   --   &   --   &   --   &   --   &     0.03 &   0.05 &   0.00 &   0.00 & 2  \\
Zr &   --   &   --   &   --   &   --   &     0.09 &   0.00 &   0.00 &   0.00 & 2  \\
Ba &  0.112 &  0.065 &  0.007 &  0.012 &     0.05 &   0.03 &   0.00 &   0.01 & 1  \\
Eu &   --   &   --   &   --   &   --   &     0.02 &   0.04 &   0.00 &   0.00 & 1  \\
\hline
\hline
 \end{tabular}
 \end{table*}

\subsubsection{NLTE effects}
Our work is based on abundances obtained under the assumption of one-dimensional (1D) model 
atmospheres and that LTE holds for all species. However, it is known that metal-poor stars, especially those with higher temperatures, can be significantly affected by NLTE and 3D effects, especially when deriving atmospheric parameters using low excitation iron lines \citep{2011A&A...528A..87M, Bergemann2012, Hansen2013}.
This prompted us to determine a photometric temperature for \cs\ .  However,  iron abundances need to be derived from the spectrum and can have NLTE corrections.
For warm BMP stars (T=6500K) with metallicities around $-2$ the corrections tend to be less than 0.1\,dex \citep{Lind2013}, which is similar to our derived Fe uncertainty (\sect{uncertainties}).

Since HD84937 has a slightly higher temperature and otherwise similar stellar parameters compared to \cs. We use the NLTE correction for the former star to estimate an order of magnitude of the NLTE corrections. 
Starting with the lightest element Li, 3D and NLTE corrections typically amount to around 0.1\,dex or less \citep{2014A&A...565A.121D, Sbordone2010, Lind2009}. The corrections for Li are less than the $\sim$0.2 dex required to lower the Li amount to the Spite plateau value.  \cite{Sbordone2010}  investigated differences in NLTE vs. LTE Li abundances using two independent model atoms \citep{Carlsson1994, Lind2009}. In both cases, stars with temperatures hotter than 6500',K have NLTE abundances that are $\sim$0.04\,dex lower than the LTE abundances. We note that for \cd\ the difference due to NLTE reported in \citep{Melendez2010} is 0.08\,dex. Regardless of 3D, NLTE corrections both BMP stars will remain above the Spite plateu.

\citet{2014A&A...565A.121D} investigated the NLTE and 3D effects on sodium and oxygen in main-sequence stars and  concluded that the departures from LTE play a dominant role at the 0.35\,dex level for Na and 0.20\,dex for oxygen. However, due to the much lower influence of convection in the atmospheres of turn-off stars 
the abundance corrections are significantly smaller in these stars -- 0.02\,dex for Na and 0.05\,dex for O. In comparison \cite{Amarsi2016} report a combined 3D, NLTE correction for O from the 6300\AA\, forbidden line of $-0.1$\,dex in a 6500\,K turn-off star. In another study of the NLTE effects of the Na D lines, which are also used in the present work, \cite{Lind2011} showed that the corrections are significant for turn-off metal-poor stars, but they slightly decrease towards hotter temperatures.

\citet{2008A&A...481..481A} determined Al NLTE corrections in metal-poor dwarfs and showed that they can reach up to 0.7\,dex for a star with parameters similar to \cs. This brings our seemingly low abundances much better in line with the approximately Solar [Al/Fe] ratios in the metal-poor halo.

For Mg, Si, Ca, the corrections are very small, on the order of 0.02\,dex or less, but at $\sim$0.25\,dex, for Cr the NLTE effects in metal-poor stars are more significant \citep{2010AA...522A...9B}. The need for large NLTE correction for Cr was recently reduced by using improved atomic data for the Cr I and Cr II lines \citep{Sneden2016}. 

Manganese NLTE effects can also be significant for metal-poor stars \citep{2008A&A...492..823B}. In the recent study of HD 84937 \cite{Sneden2016} discussed the Mn resonance lines $\lambda\lambda$ 4030, 4033, 4034\,\AA, which have corrections of more than 0.5\,dex. 
For the s-process element Sr, the detailed spectroscopic analysis of \cite{Hansen2013} 
 showed that differences between NLTE and LTE in Sr II abundances (from the 4077\,\AA-line) in the warmest, metal-poor dwarfs of their sample could reach 0.12 dex. 
 In contrast, the neutral 4607\,\AA-line experiences a much larger departure from LTE. For Ba, the study of \cite{2009A&A...494.1083A} and \cite{2015A&A...581A..70K} 
 show that, { as for Sr, NLTE will as a function of temperature act} differently on different transitions. For instance, the line at 4554\,\AA used in our study, has NLTE effects 
on the order of 0.1\,dex in BMP stars. 

Unfortunately there have been no studies on NLTE corrections spanning the large parameter range of blue stragglers or BMP stars in the literature. As we have seen above, most of the corrections remain minor (within the derived uncertainty) except from an element like Al.
Therefore, in the next section, we limit ourselves to comparing the chemical abundances of \cs\ and \cd\ to other warm dwarf stars, which have also been analysed assuming 1D-LTE. Although the chemical abundances might be subject to some corrections, the {\em relative} comparison between the abundances in stars with similar stellar parameters still sheds light on the nature of BMP stars.

\section{Discussion of BMP formation sites}\label{discussion}
\subsection{Correlation with stellar rotation}\label{Lidip}
\citet{Boesgaard1986} noted a strong decline in the Li-abundance within a very narrow temperature window $\pm$300\,K
in Pop I and old, open cluster stars. Reasons for this comprise differential rotation, meridional circulation, convection and/or diffusion effects, 
all of which involves transport of the surface Li to deeper, hotter layers, where it is easily destroyed. A trend 
of Li abundance and rotational properties thus arises naturally. 
Such a dip has not been seen in young star clusters, while it is more evident in old, metal-poor systems. 
\citet{Dearborn1992} suggest that mass loss during the main sequence phase can cause the Li-dip, which predicts that a similar 
depletion should be observed in metal-poor Pop II systems, as indeed first evidenced in \citet{Asplund2006,Bonifacio2007}. { Here we do not find evidence of a Li-dip in the range 6400--6800\,K since both single stars and binary candidates remain at the Spite Plateau, except from the BMP binaries that are naturally depleted in Li (see Fig.~\ref{Li_T_Fe}).}

As a consequence of the well-established relation between rotation and spectral type \citep[e.g.,][]{2013A&A...557L..10N}, we show in Fig.~\ref{profiles} 
selected spectral lines in \object{HD 142575} compared to their counterparts in \object{CS 22874$-$042} and \object{CD $-$48 2445}. \object{HD 142575} is also a BMP star, but it is hotter and rotates faster, as can be seen from the broadened spectral lines. For this star, \cite{Carney2005} list an A(Li)  of 1.45 dex and an EW of 2.1 m\AA\, was reported by \citet{Fulbright2000}. Our spectrum of higher SNR and higher resolution did not permit us to measure an EW for this line, which implies that HD 142575 is clearly Li-depleted. Using a sample of { halo stars that experienced mass transfer} \citet{Masseron2012} showed that all stars with large rotational velocities ($>$ 8 km\,s$^{-1}$) had depleted Li, while all stars with normal Li were slow rotators. This is confirmed by our sample, where the fast rotator HD 142575 is depleted in Li, while the stars with lower v$\,\sin\,i$  (\cs\ and \cd) have high-to-normal Li. { We note that despite the lower v$\,\sin\,i$/v$_{macro}$ \cd\ may still have experienced mass transfer (see below).}

\subsection{$s$-process and AGB enhancement}\label{sect:AGB}
We find that \cd\ is s-process enhanced (notably in La, Zr, and Sr) with respect to \cs. While both stars have a comparable abundance pattern of iron-peak and $\alpha-$elements, they show a  clear difference in their enrichment patterns at a 0.2 dex level or more in the heavy $r-$ and $s-$elements (see \tab{tab:abun}). 
 
{ An} apparent RV variation of \cd\ (seen over 40 years), and this star falling below the colour cut $V-Ks_0=1.08$ would point towards binarity and mass transfer from an AGB star affecting its abundance pattern. 
Owing to the Ba and especially La enrichment, the AGB polluting \cd\  was likely of a lower mass, since already a $2~\mathrm{M}_{\odot}$ AGB would produce and transfer much larger abundances than what we detect observationally 
(\cite{2009A&A...508.1359I} and \cite{Cristallo2011}).
That is,  a 1.3\,M$_{\odot}$ AGB or even slightly lower mass would agree better with the observed (low) Ba abundance \citep[see e.g.][]{Cristallo2011}. 
The enhancement of $s-$ process   
with respect to \cs\ suggest that the variation of RV is real. 

\subsection{Origin of BMP stars and comparison with the literature}
\cs\ shows evidence of being a regular halo star in many regards, such as its elevated [$\alpha$/Fe] ratios, which, at its metallicity of $-1.9$\,dex, 
would argue against an accretion origin from a dwarf galaxy unless the stripped 
satellite had been massive with a high star formation rate to reach this level of $\alpha$
enhancement \citep[e.g.,][]{2014ApJ...785..102H}.
In either case its abundance pattern points to a scenario where this star has had time to be
enriched by several SNe at a [Fe/H]$>-2$ dex and a mean  [$\alpha$/Fe] of 0.35\,dex.

However, at [Fe/H]$\sim -2$ (after less than 1\,Gyr on the enrichment clock), yields  from SN Ia have not appeared yet in a standard Galactic chemical evolution scenario, 
so several SNe II will have had time to explode and enrich the ISM.  

 To aid our interpretation on the BMP star formation mechanisms, we compare our abundance measurements of \cs\ and \cd\ to three other hot dwarfs from the literature. These are: 
 \object{HD~106038} (this halo star has a very similar enhancement of Li to \cs; \citet{Asplund2006}); 
 \object{HD~84937} (usually treated as a typical halo star with no signatures of binarity, pollution or extra mixing); and 
 \object{CS~22950$-$173} (with similar temperatures to our sample but normal Li abundance; \citet{Sbordone2010}).  
 
Here, \object{CS~22950$-$173} is interesting since it also belongs to the RV constant sample from PS00. Our spectrum for this star was too noisy to detect Li, but the UVES spectrum used by \cite{Sbordone2010} allowed them to  robustly detect Li. Moreover, it is important to note that \cite{Sbordone2010} determined the temperature for this metal-poor dwarf using different methods. With their  NLTE calculations for the Balmer lines they obtained a temperature for this BMP star significantly lower (6335 -- 6506\,K) than in PS00 (6800\,K), similar to what we found for \cs. 
For our comparison, we employ literature studies preferentially using LTE for the sake of consistency. 
The respective results from the literature are indicated in \tab{tab:lit}. 
 \begin{table*}
 \caption{Literature abundance for our comparison stars. }
 \label{tab:lit}
 \centering
\begin{tabular}{c | c c  | c c  | cc }
\hline
\hline
Element	&	\multicolumn{2}{c}{HD 106038}	&	\multicolumn{2}{c}{HD 84937}	&	\multicolumn{2}{c}{CS 22950-173}	\\
	&	[X/Fe]	&	Reference	&	[X/Fe]	& Reference	&	[X/Fe]	&	Reference	\\
\hline													
Li	&	2.48	&	A06	&	2.30	&	T94	&	2.20--2.30 	&	S10	\\
Mg	&	0.36	&	N97	&	0.27	&	J15	&	0.38	&	PS00	\\
Al	&	-	&		&	$-0.89$	&	M08	&	-	&		\\
Si	&	0.57	&	N97	&	0.30	&	J15	&	\dots	&	\dots	\\
Ca	&	0.21	&	N97	&	0.36	&	J15	&	0.18	&	PS00	\\
Sc	&	0.17	&	B15	&	0.13	&	J15	&	\llap{$-$}0.01	&	PS00	\\
Ti	&	0.19	&	N97	&	0.37	&	J15	&	0.53	&	PS00	\\
V	&	\dots 	&	\dots	&	0.25	&	S16	&	\dots	&	\dots	\\
Cr	&	0.02	&	N97	&	\llap{$-$}0.20	&	J15	&	\llap{$-$}0.08	&	PS00	\\
Mn	&	\llap{$-$}0.20	&	P13	&	\llap{$-$}0.27	&	S16	&	\llap{$< -$}0.13	&	PS00	\\
Co	&	0.10	&	P13	&	0.14	&	S16	&	\dots	&	\dots	\\
Ni	&	0.18	&	N97	&	0.00	&	J15	&	\dots	&	\dots	\\
Sr	&	0.56	&	H12	&	\llap{$-$}0.03 	&	M08	&	\llap{$-$}0.74	&	PS00	\\
Y	&	0.54	&	H12	&	0.00	&	M08	&	\dots	&	\dots	\\
Zr	&	0.68	&	H12	&	0.25&	M08	&	\dots	&	\dots	\\
Ba	&	0.76	&	H12	&	\llap{$-$}0.15	&	M08	&	\llap{$-$}0.04	&	PS00	\\
Eu	&	0.45	&	H12	&	0.54 	&	M08	&	\dots	&	\dots	\\												
\hline
\hline
 \end{tabular}
 \tablefoot{References:  A06: \citet{Asplund2006}, 												 
 N97: \citet{1997A&A...326..751N},	 											
B15: \citet{2015A&A...577A...9B}, 												
P13: \citet{2013ApJ...768L..13P},	 											
H13: \citet{Hansen2013},
H12: \citet{Hansen2012},	 										
T94: \citet{1994ApJ...421..318T}, 												
J15: \citet{2015A&A...582A..81J},												
S16: \citet{Sneden2016},											
M07: \citet{2007ARep...51..903M}, 												
F00: \citet{Fulbright2000}, 												
M08: \citet{Mashonkina2008},   
S10: \citet{Sbordone2010}, 											
PS00: \citet{ps00}, 												
S14: \citet{2014A&A...571A..40S}. 	}
 \end{table*}
\begin{figure*}
\hspace{-1.0cm}
\includegraphics[scale=0.75]{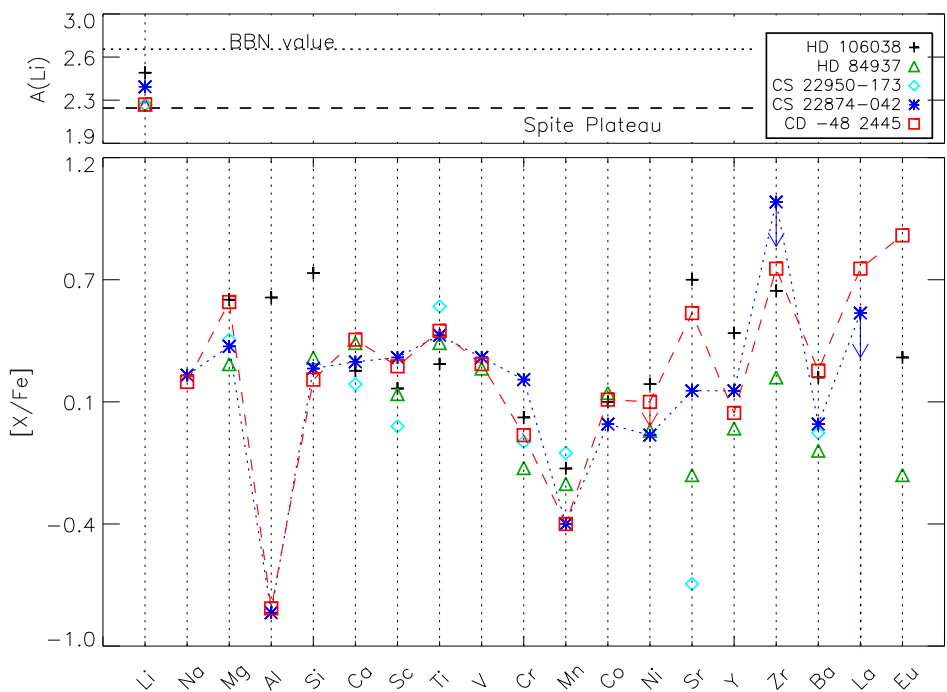}
\centering
\caption{Abundance pattern for \cs, \cd\ and our comparison stars \object{HD 106038}, \object{HD 84937}, and \object{CS 22950-173}. The abundances of the comparison stars have been taken from the literature (see \tab{tab:lit}). 
{ We also indicate the BBN Li level from WMAP \citep[A(Li)=2.7][]{Cyburt2008} as well as the Spite Plateau (2.2).}}
\label{abundance_patterns}
\end{figure*}

As it transpires, all stars have relatively homogeneous values in the $\alpha-$ and iron-peak element abundances. Their scatter can be attributed to differences in, e.g., 
different techniques for the abundance determination between the sources and differences in stellar parameters. 
This can induce a scatter in the abundances of more than 0.2\,dex, especially for those elements that are affected by hyperfine structure, such as the odd-Z iron-peak elements \citep{2014A&A...570A.122S, 2015A&A...582A..81J}.  
Other abundance peculiarities, such as the strong Si- and $s$-process enhancement in \object{HD~106038} remain hard to explain to date, even more so given the RV-constancy \citep{1997A&A...326..751N} and our colour method supports that this star is single, since it is located above the $V-Ks_0=1.08$ cut. 

It is interesting to note that the neutron-capture elements differ by more than 0.2\,dex and as such the difference are unlikely to be attributed to different measurement techniques as above
(cf. \tab{tab:abu_fin}).  In Galactic chemical evolution this is expected, where abundances of these elements { present a scatter} of up to one~dex at these metallicities \citep{2008ARA&A..46..241S, Hansen2012,2014ApJ...797..123H}.
In this regard,  \object{HD~106038} is interesting in that it is systematically enhanced in all neutron-capture elements with respect to the other stars. \citet{Smiljanic2008} suggested that a hypernova could have caused the enrichment seen in Be and Li, however, they had problems explaining the observed $^6$Li as well as other elemental abundances. As this star has a metallicity of $-1.48$ both supernovae (SN) and AGB stars can easily have contributed to the gas mixture that created this star. We therefore suggest that the observationally derived abundances are a result of AGB plus either magneto-hydrodynamic (MHD) SN \citep{Winteler2012} or neutron star mergers (NSM) which could explain the enhanced levels of the neutron-capture elements. This agrees well with a [Ba/Eu]\footnote{[Ba/Eu]$<-0.77$ indicates a pure $r-$process origin.} = 0.31 pointing towards a mixed $r-$ and $s-$process origin (\cite{Hansen2012,Mashonkina2008}, and \cite{1999ApJ...525..886A,2014ApJ...787...10B}). 

In \object{HD 84937} a very different chemical pattern is seen for the heavy elements (see Table~\ref{tab:lit} and Fig.~\ref{abundance_patterns}). Here the [Ba/Eu]$\sim-0.7$ indicates an almost pure r-process enrichment at a lower metallicity \citep[$-2.15$][]{Mashonkina2008}. This would indicate a large contribution from either MHD SN or NSM. 

The last star taken from the literature is \object{CS~22950$-$173} which is underabundant in the two known heavy elements Sr and Ba. Based on the low levels, a possible formation site could be the fast rotating massive stars \citep{Frischknecht2012}, however, the reverse abundance ratios would be expected for Sr and Ba (i.e., Sr$>$Ba). According to Arlandini et al. 1999 and \citet{Mashonkina2008} a [Sr/Ba]$<$0.62 could also indicate an $r-$process origin, which in this star would be very diluted (or inefficient { thereby making `normal'} SN an option).

Although \cs\ and \cd\ have very similar iron-peak and $\alpha-$abundances, the abundances of $s-$process elements differ. We therefore believe that \cd\ is a result of AGB yields mixed with an efficient $r-$site like NSM or MHD SN, while \cs\ is more likely enriched by massive rotating stars or normal (massive) supernovae creating the high level of $\alpha-$elements and the low level of neutron-capture elements. We note that the low Ba abundance is a bit puzzling, and  while the Ba/Eu-ratio would indicate a very r-process dominated gas origin { of \cd\,} the La/Eu-ratio would point towards a more convolved $r-$ and $s-$process mixture, which we support based on the high level of Sr and Zr in this star.

\section{Summary and Conclusions}\label{fin}
In the era of large surveys we are bound to observe blue metal-poor stars in large numbers as they are often bright. Thus, it is becoming increasingly important to understand their origin in order to explore their potential as chemical tracers and segregators of galaxy formation scenarios.
 
Blue metal-poor stars are important probes of Galactic evolution because they are either younger than typical turn-off stars of the halo, or because they are field blue stragglers (binaries). 
Here we developed a way, { which to first order} relies on colour and photometry, to separate the binaries from the single stars, immediately allowing the blue stragglers to be singled out amongst the blue stars. We find a cut at $V-Ks_0=1.08$\,mag below which the binaries fall in 97\% of the cases (including some single star pollution possibly caused by fast rotating stars). Above this cut (almost only) single stars are found. The method is further improved by { combining photometry and spectroscopy (comparing the $V-Ks_0$ to the stellar Li abundance,} a useful indicator as mentioned by \cite{Carney2005}). However, this second step requires high-resolution, high SNR spectra.
Spectral analysis of these objects is thus a vital tool for disentangling details about their origin as well as
chemical tagging of their composition and assessment of birthplace and nuclear formation processes. Further information can be drawn from mass-transfer signatures (e.g., $s-$process elements) that might have affected their evolution.  
These analyses  are, however, challenging, as their hot and metal-poor nature makes the identification of spectral lines difficult, and large telescopes are needed to obtain a spectrum with high 
enough SNR. { Moreover, repeat observations covering an extended timespan of several years are needed to assess binarity based on RV measurements in such stars.} For this reason, there is little information about their chemical composition available in the literature.  
Here we contributed to this field with a new way of separating binary from single blue stars combined detailed chemical analysis of two BMP stars in a comparison with similar stars from the literature. 
Our results can be summarised as following:
\begin{itemize}
\setlength{\itemsep}{1ex}
\item {\it Colour separation of binary and single stars:} Here we used $V-Ks_0$ vs $B-V_0$ or metallicity to separate blue binary stars from single stars. A colour cut at $V-Ks_0=1.08$\,mag was discovered. This is useful in surveys (such as Gaia) that provide colour and metallicity. Moreover, we show how Li abundances combined with photometry allow for a smooth
and fast separation of which BMP stars are binaries and which are single stars.

\item {\it Stellar parameters:}  With our improved high-quality spectrum and a new parameter determination, we found a significantly lower temperature of \cs\ than had originally been determined in PS00.
Such discrepancies have also been reported in the literature \citep{Sbordone2010}. Since our new effective temperature lowered the Li abundance by almost one dex, this emphasizes the need
for accurate and precise stellar parameters when assessing the chemical nature of this class of stars. 
\item {\it The behaviour of Li in blue stragglers}: 
It has been suggested that BMP stars with a regular plateau value of A(Li) should be intermediate-age stars, while Li-depleted stars are more likely to be blue stragglers. Our work showed that this distinction is not that straightforward. 
One decisive factor is stellar mass, as several theoretical and observational studies have ascertained that the depletion from the primordial BBN 
value acts more efficiently in lower-mass stars through mass-dependent diffusion 
\citep[e.g.][]{Richard2005,Gonzalez2008,Melendez2010,Aoki2012}. Similarly, if mass transfer occurred in these systems, diffusion can be inhibited such that Li could retain its original (high) levels. This is bolstered by our finding of the Li-plateau value in \cd\, which shows other signs of mass transfer; the fact that the Li-enhanced star in our study, \cs\, is clearly a single star, gives weight to our proposed separation. See also \cite{Jofre2016}  for discussion on possible Li enhancement in evolved blue stragglers that are not in binary systems.
\item {\it Lithium and rotation in RV constant stars:} \object{HD 142575} is found to be Li depleted and a fast rotating star. \cs\  on the other hand is slowly rotating and Li-enriched confirming that there is a correlation between the rotation and the Li abundance.
This agrees with the results of \cite{Masseron2012}.
\item {\it RV constancy of \cs\ over 22 years}: Our new RV { determination 15 years after} the last RV measurement of PS00 confirms that \cs\ does not have a companion that we can detect { neither in RV nor in colour}.
\item {\it Origin of BMP stars}: From its abundances of $r-$ and $s-$process elements, we conclude that \cd\  might have received prenatal gas from a NSM and is a possible blue straggler that accreted mass from an AGB. It may still have a binary companion in the form of a white dwarf. \cs\ on the other hand has experienced a very different enrichment, possibly from massive stars with inefficient heavy element production. Both stars have a relatively high $\alpha$-to-Fe ratio making them in situ formed Milky Way stars.

\end{itemize}
More high SNR spectra and accurate photometry for BMP stars are needed to revise their stellar parameters, determine { elemental abundances}, and gain extra RV measurement that will give us more insights { into} how various subgroups of these blue stars are created. Additionally, the photometry will help confirm and generalise the colour cut discovered.
The Li, $\alpha$, $r-$ and $s-$element abundances allow us to distinguish the BMP fraction that are binaries and suffered from mass transfer from the ones that are accreted early on in a galaxy merger event. Hence abundance studies of these elements in such stars will help enlightning us on galaxy formation and evolution and this has an important outlook for the future large surveys that are efficient in the red-wavelenght range (e.g., 4MOST, HIRES, WEAVE).

\begin{acknowledgements}  
{ We thank the anonymous referee for helpful and constructive input and G. W. Preston for fruitful discussions.}
CJH thanks R. Izzard for a useful discussion and acknowledges support from the research grant VKR023371 by the VILLUM FOUNDATION.
PJ dedicates this work to J. Jofr\'e, as this analysis was not only motivated to study lithium in BMP stars, but also to find peace in a difficult moment of severe illness.  This work has been partially supported by ERC grant number 320360. PJ also acknowledges T. Masseron and C. Tout for useful discussions on this subject, as well as King's College Cambridge for partially supporting this research. 
\end{acknowledgements}

\small

\newpage
\onecolumn
\begin{appendix}
\section{Line list}
\begin{longtable}{c c c c c}
\caption{Line list used in our analysis. References correspond to WIE: \citet{1996atpc.book.....W}; SNE: \cite{2014ApJS..214...26S}; LAW: \cite{2001ApJ...556..452L,2001ApJ...563.1075L}; SOB: \cite{2007ApJ...667.1267S}; YAN: \cite{1998PhRvA..57.1652Y}; GAL: \cite{2012A&A...538A.118G}. \label{info_lines}}\\
\hline
\hline
Wavelength [\AA]& Species & $\log gf$ &E.P. [eV] & Reference  \\
\hline
\endfirsthead
\multicolumn{5}{c}%
{\tablename\ \thetable\ -- \textit{Continued}} \\
\hline
Wavelength [\AA]& Species & $\log gf$ &E.P. [eV] & Reference  \\
\hline
\endhead
Li& & & & \\
 6707.764   & 3.0 &   -0.002  &    0.000  &  YAN   \\
 6707.915   & 3.0 &   -0.303  &    0.000  &  YAN   \\
 6707.921   & 3.0 &   -0.002  &    0.000  &  YAN  \\
 6708.072   & 3.0 &   -0.303  &    0.000  &  YAN   \\
\hline 
O& & & & \\
7771.940  &   8.0  &    0.370  &    9.140  &  WIE   \\
7774.170  &   8.0  &    0.220  &    9.140  &  WIE   \\
7775.390  &   8.0  &    0.000  &    9.140  &  WIE   \\
\hline 
Na& & & & \\
5889.951  &  11.0  &    0.117  &    0.000  &  SNE   \\
5895.924  &  11.0  &   -0.184  &    0.000  &  SNE   \\
\hline 
Mg& & & & \\
4057.505  &  12.0  &   -1.201  &    4.343  &  SNE   \\
4167.271  &  12.0  &   -1.004  &    4.343  &  SNE   \\
4702.991  &  12.0  &   -0.666  &    4.343  &  SNE   \\
5528.405  &  12.0  &   -0.620  &    4.343  &  SNE   \\
\hline 
Al& & & & \\
3944.006  &  13.0  &   -0.623  &    0.000  &  SNE   \\
3961.520  &  13.0  &   -0.323  &    0.014  &  SNE   \\
\hline 
Si& & & & \\
3905.523  &  14.0  &   -1.090  &    1.907  &  SNE   \\
7405.772  &  14.0  &   -0.820  &    5.609  &  SNE   \\
7415.948  &  14.0  &   -0.500  &    5.611  &  SNE   \\
\hline 
Ca& & & & \\
4226.728  &  20.0  &    0.243  &    0.000  &  SNE   \\
4283.011  &  20.0  &   -0.224  &    1.884  &  SNE   \\
4289.367  &  20.0  &   -0.303  &    1.878  &  SNE   \\
4298.988  &  20.0  &   -0.412  &    1.884  &  SNE   \\
4302.528  &  20.0  &    0.275  &    1.897  &  SNE   \\
4307.744  &  20.0  &   -0.256  &    1.884  &  SNE   \\
4318.652  &  20.0  &   -0.208  &    1.897  &  SNE   \\
4425.437  &  20.0  &   -0.385  &    1.878  &  SNE   \\
4434.957  &  20.0  &   -0.029  &    1.884  &  SNE   \\
4435.679  &  20.0  &   -0.500  &    1.884  &  SNE   \\
4454.779  &  20.0  &    0.252  &    1.897  &  SNE   \\
4455.887  &  20.0  &   -0.510  &    1.897  &  SNE   \\
4585.865  &  20.0  &   -0.186  &    2.524  &  SNE   \\
5264.237  &  20.0  &   -0.720  &    2.521  &  SNE   \\
5265.556  &  20.0  &   -0.260  &    2.521  &  SNE   \\
5588.749  &  20.0  &    0.210  &    2.524  &  SNE   \\
5594.462  &  20.0  &   -0.050  &    2.521  &  SNE   \\
5598.480  &  20.0  &   -0.220  &    2.519  &  SNE   \\
5601.277  &  20.0  &   -0.690  &    2.524  &  SNE   \\
5857.451  &  20.0  &    0.230  &    2.930  &  SNE   \\
6102.723  &  20.0  &   -0.890  &    1.878  &  SNE   \\
6122.217  &  20.0  &   -0.409  &    1.884  &  SNE   \\
6162.173  &  20.0  &    0.100  &    1.897  &  SNE   \\
6439.075  &  20.0  &    0.470  &    2.524  &  SNE   \\
6462.567  &  20.0  &    0.310  &    2.521  &  SNE   \\
6493.781  &  20.0  &    0.140  &    2.519  &  SNE   \\
\hline 
Sc& & & & \\
 4246.813   & 21.1 &   -0.385  &    0.315  &  LAW   \\
 4246.813   & 21.1 &   -0.955  &    0.315  &  LAW   \\
 4246.820   & 21.1 &   -0.783  &    0.315  &  LAW   \\
 4246.820   & 21.1 &   -0.797  &    0.315  &  LAW   \\
 4246.820   & 21.1 &   -0.955  &    0.315  &  LAW   \\
 4246.826   & 21.1 &   -0.797  &    0.315  &  LAW   \\
 4246.826   & 21.1 &   -0.808  &    0.315  &  LAW   \\
 4246.826   & 21.1 &   -1.479  &    0.315  &  LAW   \\
 4246.831   & 21.1 &   -0.808  &    0.315  &  LAW   \\
 4246.831   & 21.1 &   -0.981  &    0.315  &  LAW   \\
 4246.831   & 21.1 &   -2.905  &    0.315  &  LAW   \\
 4246.834   & 21.1 &   -0.981  &    0.315  &  LAW   \\
 4246.834   & 21.1 &   -1.157  &    0.315  &  LAW   \\
 4314.076   & 21.1 &   -2.782  &    0.618  &  LAW   \\
 4314.078   & 21.1 &   -1.569  &    0.618  &  LAW   \\
 4314.079   & 21.1 &   -2.379  &    0.618  &  LAW   \\
 4314.080   & 21.1 &   -0.640  &    0.618  &  LAW   \\
 4314.081   & 21.1 &   -1.364  &    0.618  &  LAW   \\
 4314.082   & 21.1 &   -0.761  &    0.618  &  LAW   \\
 4314.082   & 21.1 &   -2.167  &    0.618  &  LAW   \\
 4314.083   & 21.1 &   -1.298  &    0.618  &  LAW   \\
 4314.083   & 21.1 &   -2.057  &    0.618  &  LAW   \\
 4314.084   & 21.1 &   -0.897  &    0.618  &  LAW   \\
 4314.084   & 21.1 &   -1.302  &    0.618  &  LAW   \\
 4314.085   & 21.1 &   -1.053  &    0.618  &  LAW   \\
 4314.085   & 21.1 &   -2.036  &    0.618  &  LAW   \\
 4314.086   & 21.1 &   -1.238  &    0.618  &  LAW   \\
 4314.086   & 21.1 &   -1.363  &    0.618  &  LAW   \\
 4314.086   & 21.1 &   -1.485  &    0.618  &  LAW   \\
 4314.086   & 21.1 &   -2.145  &    0.618  &  LAW   \\
 4314.087   & 21.1 &   -1.472  &    0.618  &  LAW   \\
 4314.087   & 21.1 &   -1.668  &    0.618  &  LAW   \\
 4314.087   & 21.1 &   -1.814  &    0.618  &  LAW   \\
 4320.725   & 21.1 &   -2.448  &    0.605  &  LAW   \\
 4320.727   & 21.1 &   -2.030  &    0.605  &  LAW   \\
 4320.728   & 21.1 &   -1.650  &    0.605  &  LAW   \\
 4320.728   & 21.1 &   -1.796  &    0.605  &  LAW   \\
 4320.729   & 21.1 &   -1.448  &    0.605  &  LAW   \\
 4320.729   & 21.1 &   -1.553  &    0.605  &  LAW   \\
 4320.730   & 21.1 &   -1.232  &    0.605  &  LAW   \\
 4320.730   & 21.1 &   -1.265  &    0.605  &  LAW   \\
 4320.730   & 21.1 &   -1.291  &    0.605  &  LAW   \\
 4320.730   & 21.1 &   -1.474  &    0.605  &  LAW   \\
 4320.732   & 21.1 &   -1.775  &    0.605  &  LAW   \\
 4320.733   & 21.1 &   -1.086  &    0.605  &  LAW   \\
 4320.733   & 21.1 &   -1.357  &    0.605  &  LAW   \\
 4320.734   & 21.1 &   -0.708  &    0.605  &  LAW   \\
 4320.734   & 21.1 &   -0.879  &    0.605  &  LAW   \\
 4324.984   & 21.1 &   -1.931  &    0.595  &  LAW  \\
 4324.986   & 21.1 &   -1.491  &    0.595  &  LAW  \\
 4324.988   & 21.1 &   -1.220  &    0.595  &  LAW  \\
 4324.992   & 21.1 &   -1.116  &    0.595  &  LAW  \\
 4324.992   & 21.1 &   -1.232  &    0.595  &  LAW  \\
 4324.993   & 21.1 &   -1.236  &    0.595  &  LAW  \\
 4324.999   & 21.1 &   -1.491  &    0.595  &  LAW  \\
 4325.000   & 21.1 &   -1.036  &    0.595  &  LAW  \\
 4325.002   & 21.1 &   -0.743  &    0.595  &  LAW  \\
 4400.379   & 21.1 &   -2.010  &    0.605  &  LAW  \\
 4400.383   & 21.1 &   -1.202  &    0.605  &  LAW  \\
 4400.383   & 21.1 &   -1.815  &    0.605  &  LAW  \\
 4400.387   & 21.1 &   -1.430  &    0.605  &  LAW  \\
 4400.387   & 21.1 &   -1.762  &    0.605  &  LAW  \\
 4400.390   & 21.1 &   -1.716  &    0.605  &  LAW  \\
 4400.390   & 21.1 &   -1.788  &    0.605  &  LAW  \\
 4400.392   & 21.1 &   -2.010  &    0.605  &  LAW  \\
 4400.393   & 21.1 &   -1.887  &    0.605  &  LAW  \\
 4400.393   & 21.1 &   -2.102  &    0.605  &  LAW  \\
 4400.394   & 21.1 &   -1.815  &    0.605  &  LAW  \\
 4400.395   & 21.1 &   -2.732  &    0.605  &  LAW  \\
 4400.396   & 21.1 &   -1.762  &    0.605  &  LAW  \\
 4400.396   & 21.1 &   -2.109  &    0.605  &  LAW  \\
 4400.398   & 21.1 &   -1.788  &    0.605  &  LAW  \\
 4400.398   & 21.1 &   -1.887  &    0.605  &  LAW  \\
 4400.398   & 21.1 &   -2.586  &    0.605  &  LAW  \\
 4400.399   & 21.1 &   -2.109  &    0.605  &  LAW  \\
 4415.543   & 21.1 &   -1.864  &    0.595  &  LAW  \\
 4415.548   & 21.1 &   -1.706  &    0.595  &  LAW  \\
 4415.552   & 21.1 &   -1.717  &    0.595  &  LAW  \\
 4415.554   & 21.1 &   -1.294  &    0.595  &  LAW  \\
 4415.556   & 21.1 &   -1.692  &    0.595  &  LAW  \\
 4415.556   & 21.1 &   -1.890  &    0.595  &  LAW  \\
 4415.559   & 21.1 &   -2.388  &    0.595  &  LAW  \\
 4415.560   & 21.1 &   -3.814  &    0.595  &  LAW  \\
 4415.562   & 21.1 &   -2.066  &    0.595  &  LAW  \\
 4415.567   & 21.1 &   -1.706  &    0.595  &  LAW  \\
 4415.567   & 21.1 &   -1.717  &    0.595  &  LAW  \\
 4415.567   & 21.1 &   -1.864  &    0.595  &  LAW  \\
 4415.567   & 21.1 &   -1.890  &    0.595  &  LAW  \\
\hline 
Ti& & & & \\
3491.049  &  22.1  &   -1.100  &    0.112  &  LAW   \\
3504.891  &  22.1  &    0.380  &    1.890  &  LAW   \\
3535.407  &  22.1  &    0.010  &    2.060  &  LAW   \\
3596.047  &  22.1  &   -1.070  &    0.607  &  LAW   \\
3659.761  &  22.1  &   -0.540  &    1.581  &  LAW   \\
3662.232  &  22.1  &   -0.540  &    1.565  &  LAW   \\
3741.638  &  22.1  &   -0.070  &    1.581  &  LAW   \\
3761.321  &  22.1  &    0.180  &    0.573  &  LAW   \\
3813.388  &  22.1  &   -1.890  &    0.607  &  LAW   \\
3814.580  &  22.1  &   -1.680  &    0.573  &  LAW   \\
3913.461  &  22.1  &   -0.360  &    1.115  &  LAW   \\
3948.671  &  22.0  &   -0.400  &    0.000  &  LAW   \\
3981.762  &  22.0  &   -0.270  &    0.000  &  LAW   \\
3989.758  &  22.0  &   -0.130  &    0.021  &  LAW   \\
3998.636  &  22.0  &    0.020  &    0.048  &  LAW   \\
4012.384  &  22.1  &   -1.780  &    0.573  &  LAW   \\
4025.129  &  22.1  &   -2.110  &    0.607  &  LAW   \\
4028.338  &  22.1  &   -0.920  &    1.890  &  LAW   \\
4053.821  &  22.1  &   -1.070  &    1.891  &  LAW   \\
4161.529  &  22.1  &   -2.090  &    1.083  &  LAW   \\
4163.644  &  22.1  &   -0.130  &    2.588  &  LAW   \\
4171.904  &  22.1  &   -0.300  &    2.596  &  LAW   \\
4173.533  &  22.1  &   -1.880  &    1.083  &  LAW   \\
4290.215  &  22.1  &   -0.870  &    1.164  &  LAW   \\
4300.042  &  22.1  &   -0.460  &    1.179  &  LAW   \\
4301.923  &  22.1  &   -1.210  &    1.160  &  LAW   \\
4305.907  &  22.0  &    0.490  &    0.848  &  LAW   \\
4312.860  &  22.1  &   -1.120  &    1.179  &  LAW   \\
4394.059  &  22.1  &   -1.770  &    1.220  &  LAW   \\
4395.031  &  22.1  &   -0.540  &    1.083  &  LAW   \\
4395.839  &  22.1  &   -1.930  &    1.242  &  LAW   \\
4399.765  &  22.1  &   -1.200  &    1.236  &  LAW   \\
4417.719  &  22.1  &   -1.430  &    1.164  &  SNE   \\
4443.801  &  22.1  &   -0.710  &    1.079  &  LAW   \\
4450.482  &  22.1  &   -1.520  &    1.083  &  LAW   \\
4468.493  &  22.1  &   -0.630  &    1.130  &  LAW   \\
4501.270  &  22.1  &   -0.770  &    1.115  &  LAW   \\
4529.480  &  22.1  &   -1.750  &    1.571  &  LAW   \\
4533.239  &  22.0  &    0.540  &    0.848  &  LAW   \\
4534.776  &  22.0  &    0.350  &    0.835  &  LAW   \\
4549.622  &  22.1  &   -0.220  &    1.583  &  LAW   \\
4563.761  &  22.1  &   -0.960  &    1.220  &  SNE   \\
4571.971  &  22.1  &   -0.310  &    1.571  &  LAW   \\
4805.085  &  22.1  &   -1.100  &    2.060  &  SNE   \\
4981.731  &  22.0  &    0.570  &    0.848  &  LAW   \\
5007.209  &  22.0  &    0.170  &    0.818  &  LAW   \\
5129.156  &  22.1  &   -1.340  &    1.890  &  LAW   \\
5226.543  &  22.1  &   -1.300  &    1.565  &  SNE   \\
\hline 
V& & & & \\
4005.706  &  23.1  &   -0.460  &    1.816  &  SNE    \\
\hline 
Cr& & & & \\
3919.150  &  24.0  &   -0.710  &    1.029  &  SOB   \\
4254.330  &  24.0  &   -0.090  &    0.000  &  SOB   \\
4274.800  &  24.0  &   -0.220  &    0.000  &  SOB   \\
4289.720  &  24.0  &   -0.370  &    0.000  &  SOB   \\
5206.040  &  24.0  &    0.020  &    0.941  &  SOB   \\
5208.420  &  24.0  &    0.170  &    0.941  &  SOB   \\
\hline 
Mn& & & & \\
 4030.492  & 25.0 &   -5.078  &    3.071  &  KUR  \\
 4030.495  & 25.0 &   -4.756  &    3.071  &  KUR  \\
 4030.500  & 25.0 &   -4.932  &    3.071  &  KUR  \\
 4030.511  & 25.0 &   -5.124  &    3.071  &  KUR  \\
 4030.516  & 25.0 &   -4.601  &    3.071  &  KUR  \\
 4030.524  & 25.0 &   -4.522  &    3.071  &  KUR  \\
 4030.540  & 25.0 &   -5.300  &    3.071  &  KUR  \\
 4030.548  & 25.0 &   -4.580  &    3.071  &  KUR  \\
 4030.558  & 25.0 &   -4.249  &    3.071  &  KUR  \\
 4030.580  & 25.0 &   -5.680  &    3.071  &  KUR  \\
 4030.590  & 25.0 &   -4.726  &    3.071  &  KUR  \\
 4030.602  & 25.0 &   -4.037  &    3.071  &  KUR  \\
 4030.730  & 25.0 &   -1.064  &    0.000  &  LAW  \\
 4030.744  & 25.0 &   -1.982  &    0.000  &  LAW  \\
 4030.746  & 25.0 &   -1.204  &    0.000  &  LAW  \\
 4030.756  & 25.0 &   -3.200  &    0.000  &  LAW  \\
 4030.758  & 25.0 &   -1.806  &    0.000  &  LAW  \\
 4030.759  & 25.0 &   -1.362  &    0.000  &  LAW  \\
 4030.767  & 25.0 &   -2.848  &    0.000  &  LAW  \\
 4030.769  & 25.0 &   -1.780  &    0.000  &  LAW  \\
 4030.770  & 25.0 &   -1.546  &    0.000  &  LAW  \\
 4030.775  & 25.0 &   -2.722  &    0.000  &  LAW  \\
 4030.777  & 25.0 &   -1.847  &    0.000  &  LAW  \\
 4030.777  & 25.0 &   -1.768  &    0.000  &  LAW  \\
 4030.781  & 25.0 &   -2.802  &    0.000  &  LAW  \\
 4030.782  & 25.0 &   -2.023  &    0.000  &  LAW  \\
 4030.782  & 25.0 &   -2.053  &    0.000  &  LAW  \\
 4033.044  & 25.0 &   -1.229  &    0.000  &  LAW  \\
 4033.046  & 25.0 &   -2.007  &    0.000  &  LAW  \\
 4033.056  & 25.0 &   -2.007  &    0.000  &  LAW  \\
 4033.058  & 25.0 &   -1.492  &    0.000  &  LAW  \\
 4033.060  & 25.0 &   -1.844  &    0.000  &  LAW  \\
 4033.068  & 25.0 &   -1.844  &    0.000  &  LAW  \\
 4033.070  & 25.0 &   -1.823  &    0.000  &  LAW  \\
 4033.071  & 25.0 &   -1.839  &    0.000  &  LAW  \\
 4033.078  & 25.0 &   -1.839  &    0.000  &  LAW  \\
 4033.079  & 25.0 &   -2.270  &    0.000  &  LAW  \\
 4033.080  & 25.0 &   -1.941  &    0.000  &  LAW  \\
 4033.084  & 25.0 &   -1.941  &    0.000  &  LAW  \\
 4033.085  & 25.0 &   -2.969  &    0.000  &  LAW  \\
 4033.085  & 25.0 &   -2.203  &    0.000  &  LAW  \\
 4033.088  & 25.0 &   -2.203  &    0.000  &  LAW  \\
 4033.178  & 25.0 &   -2.952  &    3.133  &  KUR  \\
 4033.178  & 25.0 &   -3.596  &    3.133  &  KUR  \\
 4033.184  & 25.0 &   -2.898  &    3.133  &  KUR  \\
 4033.184  & 25.0 &   -3.399  &    3.133  &  KUR  \\
 4033.184  & 25.0 &   -3.596  &    3.133  &  KUR  \\
 4033.193  & 25.0 &   -2.792  &    3.133  &  KUR  \\
 4033.193  & 25.0 &   -3.343  &    3.133  &  KUR  \\
 4033.193  & 25.0 &   -3.399  &    3.133  &  KUR  \\
 4033.203  & 25.0 &   -2.668  &    3.133  &  KUR  \\
 4033.203  & 25.0 &   -3.343  &    3.133  &  KUR  \\
 4033.203  & 25.0 &   -3.384  &    3.133  &  KUR  \\
 4033.216  & 25.0 &   -2.540  &    3.133  &  KUR  \\
 4033.216  & 25.0 &   -3.384  &    3.133  &  KUR  \\
 4033.216  & 25.0 &   -3.574  &    3.133  &  KUR  \\
 4033.231  & 25.0 &   -2.415  &    3.133  &  KUR  \\
 4033.231  & 25.0 &   -3.574  &    3.133  &  KUR  \\
 4034.338  & 25.0 &   -4.894  &    3.133  &  KUR  \\
 4034.338  & 25.0 &   -5.405  &    3.133  &  KUR  \\
 4034.338  & 25.0 &   -6.502  &    3.133  &  KUR  \\
 4034.345  & 25.0 &   -4.761  &    3.133  &  KUR  \\
 4034.345  & 25.0 &   -5.217  &    3.133  &  KUR  \\
 4034.345  & 25.0 &   -6.371  &    3.133  &  KUR  \\
 4034.353  & 25.0 &   -4.631  &    3.133  &  KUR  \\
 4034.353  & 25.0 &   -5.162  &    3.133  &  KUR  \\
 4034.353  & 25.0 &   -6.468  &    3.133  &  KUR  \\
 4034.364  & 25.0 &   -4.509  &    3.133  &  KUR  \\
 4034.364  & 25.0 &   -5.200  &    3.133  &  KUR  \\
 4034.364  & 25.0 &   -6.803  &    3.133  &  KUR  \\
 4034.377  & 25.0 &   -4.396  &    3.133  &  KUR  \\
 4034.377  & 25.0 &   -5.388  &    3.133  &  KUR  \\
 4034.392  & 25.0 &   -4.292  &    3.133  &  KUR  \\
 4034.469  & 25.0 &   -1.358  &    0.000  &  LAW  \\
 4034.471  & 25.0 &   -2.047  &    0.000  &  LAW  \\
 4034.472  & 25.0 &   -3.002  &    0.000  &  LAW  \\
 4034.483  & 25.0 &   -1.570  &    0.000  &  LAW  \\
 4034.485  & 25.0 &   -1.901  &    0.000  &  LAW  \\
 4034.486  & 25.0 &   -2.621  &    0.000  &  LAW  \\
 4034.494  & 25.0 &   -1.843  &    0.000  &  LAW  \\
 4034.495  & 25.0 &   -1.922  &    0.000  &  LAW  \\
 4034.496  & 25.0 &   -2.445  &    0.000  &  LAW  \\
 4034.501  & 25.0 &   -2.253  &    0.000  &  LAW  \\
 4034.502  & 25.0 &   -2.077  &    0.000  &  LAW  \\
 4034.503  & 25.0 &   -2.399  &    0.000  &  LAW  \\
 4034.717  & 25.0 &   -3.312  &    3.133  &  KUR  \\
 4034.717  & 25.0 &   -3.567  &    3.133  &  KUR  \\
 4034.722  & 25.0 &   -3.248  &    3.133  &  KUR  \\
 4034.722  & 25.0 &   -3.369  &    3.133  &  KUR  \\
 4034.722  & 25.0 &   -3.567  &    3.133  &  KUR  \\
 4034.729  & 25.0 &   -3.073  &    3.133  &  KUR  \\
 4034.729  & 25.0 &   -3.304  &    3.133  &  KUR  \\
 4034.729  & 25.0 &   -3.369  &    3.133  &  KUR  \\
 4034.738  & 25.0 &   -2.881  &    3.133  &  KUR  \\
 4034.738  & 25.0 &   -3.304  &    3.133  &  KUR  \\
 4034.738  & 25.0 &   -3.335  &    3.133  &  KUR  \\
 4034.750  & 25.0 &   -2.696  &    3.133  &  KUR  \\
 4034.750  & 25.0 &   -3.335  &    3.133  &  KUR  \\
 4034.750  & 25.0 &   -3.517  &    3.133  &  KUR  \\
 4034.765  & 25.0 &   -2.525  &    3.133  &  KUR  \\
 4034.765  & 25.0 &   -3.517  &    3.133  &  KUR  \\
\hline 
Co& & & & \\
 3872.945  & 27.0 &   -2.183  &    3.512  &  KUR  \\
 3872.945  & 27.0 &   -2.379  &    3.512  &  KUR  \\
 3872.953  & 27.0 &   -2.158  &    3.512  &  KUR  \\
 3872.953  & 27.0 &   -2.169  &    3.512  &  KUR  \\
 3872.953  & 27.0 &   -2.379  &    3.512  &  KUR  \\
 3872.965  & 27.0 &   -2.031  &    3.512  &  KUR  \\
 3872.965  & 27.0 &   -2.055  &    3.512  &  KUR  \\
 3872.965  & 27.0 &   -2.158  &    3.512  &  KUR  \\
 3872.981  & 27.0 &   -1.864  &    3.512  &  KUR  \\
 3872.981  & 27.0 &   -2.012  &    3.512  &  KUR  \\
 3872.981  & 27.0 &   -2.055  &    3.512  &  KUR  \\
 3873.001  & 27.0 &   -1.695  &    3.512  &  KUR  \\
 3873.001  & 27.0 &   -2.012  &    3.512  &  KUR  \\
 3873.001  & 27.0 &   -2.023  &    3.512  &  KUR  \\
 3873.026  & 27.0 &   -1.536  &    3.512  &  KUR  \\
 3873.026  & 27.0 &   -2.023  &    3.512  &  KUR  \\
 3873.026  & 27.0 &   -2.100  &    3.512  &  KUR  \\
 3873.054  & 27.0 &   -1.388  &    3.512  &  KUR  \\
 3873.054  & 27.0 &   -2.100  &    3.512  &  KUR  \\
 3873.054  & 27.0 &   -2.315  &    3.512  &  KUR  \\
 3873.086  & 27.0 &   -1.252  &    3.512  &  KUR  \\
 3873.086  & 27.0 &   -2.315  &    3.512  &  KUR  \\
 3873.006  & 27.0 &   -4.220  &    2.278  &  KUR  \\
 3873.006  & 27.0 &   -4.220  &    2.278  &  KUR  \\
 3873.006  & 27.0 &   -4.521  &    2.278  &  KUR  \\
 3873.010  & 27.0 &   -4.056  &    2.278  &  KUR  \\
 3873.010  & 27.0 &   -4.082  &    2.278  &  KUR  \\
 3873.010  & 27.0 &   -4.396  &    2.278  &  KUR  \\
 3873.017  & 27.0 &   -3.799  &    2.278  &  KUR  \\
 3873.017  & 27.0 &   -4.048  &    2.278  &  KUR  \\
 3873.017  & 27.0 &   -4.646  &    2.278  &  KUR  \\
 3873.026  & 27.0 &   -3.583  &    2.278  &  KUR  \\
 3873.026  & 27.0 &   -4.209  &    2.278  &  KUR  \\
 3873.026  & 27.0 &   -5.074  &    2.278  &  KUR  \\
 3873.071  & 27.0 &   -3.642  &    0.431  &  KUR  \\
 3873.073  & 27.0 &   -3.244  &    0.431  &  KUR  \\
 3873.075  & 27.0 &   -3.040  &    0.431  &  KUR  \\
 3873.077  & 27.0 &   -2.943  &    0.431  &  KUR  \\
 3873.079  & 27.0 &   -2.943  &    0.431  &  KUR  \\
 3873.082  & 27.0 &   -3.098  &    0.431  &  KUR  \\
 3873.089  & 27.0 &   -2.399  &    0.431  &  KUR  \\
 3873.090  & 27.0 &   -2.203  &    0.431  &  KUR  \\
 3873.092  & 27.0 &   -2.098  &    0.431  &  KUR  \\
 3873.093  & 27.0 &   -2.051  &    0.431  &  KUR  \\
 3873.093  & 27.0 &   -2.466  &    0.431  &  KUR  \\
 3873.095  & 27.0 &   -2.057  &    0.431  &  KUR  \\
 3873.098  & 27.0 &   -2.130  &    0.431  &  KUR  \\
 3873.098  & 27.0 &   -2.203  &    0.431  &  KUR  \\
 3873.100  & 27.0 &   -2.341  &    0.431  &  KUR  \\
 3873.103  & 27.0 &   -1.999  &    0.431  &  KUR  \\
 3873.108  & 27.0 &   -1.829  &    0.431  &  KUR  \\
 3873.114  & 27.0 &   -1.683  &    0.431  &  KUR  \\
 3873.120  & 27.0 &   -1.554  &    0.431  &  KUR  \\
 3873.127  & 27.0 &   -1.438  &    0.431  &  KUR  \\
 3873.133  & 27.0 &   -1.333  &    0.431  &  KUR  \\
 3873.919  & 27.0 &   -3.454  &    0.513  &  KUR  \\
 3873.925  & 27.0 &   -3.044  &    0.513  &  KUR  \\
 3873.930  & 27.0 &   -2.822  &    0.513  &  KUR  \\
 3873.935  & 27.0 &   -2.697  &    0.513  &  KUR  \\
 3873.939  & 27.0 &   -2.340  &    0.513  &  KUR  \\
 3873.940  & 27.0 &   -2.646  &    0.513  &  KUR  \\
 3873.942  & 27.0 &   -2.144  &    0.513  &  KUR  \\
 3873.944  & 27.0 &   -2.090  &    0.513  &  KUR  \\
 3873.944  & 27.0 &   -2.676  &    0.513  &  KUR  \\
 3873.946  & 27.0 &   -2.112  &    0.513  &  KUR  \\
 3873.947  & 27.0 &   -2.203  &    0.513  &  KUR  \\
 3873.947  & 27.0 &   -2.391  &    0.513  &  KUR  \\
 3873.954  & 27.0 &   -2.646  &    0.513  &  KUR  \\
 3873.957  & 27.0 &   -2.258  &    0.513  &  KUR  \\
 3873.959  & 27.0 &   -2.003  &    0.513  &  KUR  \\
 3873.961  & 27.0 &   -1.805  &    0.513  &  KUR  \\
 3873.962  & 27.0 &   -1.641  &    0.513  &  KUR  \\
 3873.963  & 27.0 &   -1.500  &    0.513  &  KUR  \\
 3995.269  & 27.0 &   -2.026  &    0.922  &  KUR  \\
 3995.271  & 27.0 &   -1.763  &    0.922  &  KUR  \\
 3995.271  & 27.0 &   -1.959  &    0.922  &  KUR  \\
 3995.276  & 27.0 &   -1.559  &    0.922  &  KUR  \\
 3995.276  & 27.0 &   -1.763  &    0.922  &  KUR  \\
 3995.276  & 27.0 &   -2.658  &    0.922  &  KUR  \\
 3995.282  & 27.0 &   -1.389  &    0.922  &  KUR  \\
 3995.282  & 27.0 &   -1.658  &    0.922  &  KUR  \\
 3995.282  & 27.0 &   -2.503  &    0.922  &  KUR  \\
 3995.290  & 27.0 &   -1.243  &    0.922  &  KUR  \\
 3995.290  & 27.0 &   -1.611  &    0.922  &  KUR  \\
 3995.290  & 27.0 &   -2.503  &    0.922  &  KUR  \\
 3995.301  & 27.0 &   -1.114  &    0.922  &  KUR  \\
 3995.301  & 27.0 &   -1.617  &    0.922  &  KUR  \\
 3995.301  & 27.0 &   -2.600  &    0.922  &  KUR  \\
 3995.313  & 27.0 &   -0.998  &    0.922  &  KUR  \\
 3995.313  & 27.0 &   -1.690  &    0.922  &  KUR  \\
 3995.313  & 27.0 &   -2.804  &    0.922  &  KUR  \\
 3995.328  & 27.0 &   -0.893  &    0.922  &  KUR  \\
 3995.328  & 27.0 &   -1.901  &    0.922  &  KUR  \\
 3995.328  & 27.0 &   -3.202  &    0.922  &  KUR  \\
 4121.294  & 27.0 &   -0.993  &    0.922  &  KUR  \\
 4121.301  & 27.0 &   -1.098  &    0.922  &  KUR  \\
 4121.308  & 27.0 &   -1.214  &    0.922  &  KUR  \\
 4121.313  & 27.0 &   -1.343  &    0.922  &  KUR  \\
 4121.316  & 27.0 &   -2.001  &    0.922  &  KUR  \\
 4121.318  & 27.0 &   -1.489  &    0.922  &  KUR  \\
 4121.321  & 27.0 &   -1.790  &    0.922  &  KUR  \\
 4121.322  & 27.0 &   -1.659  &    0.922  &  KUR  \\
 4121.324  & 27.0 &   -1.717  &    0.922  &  KUR  \\
 4121.326  & 27.0 &   -1.863  &    0.922  &  KUR  \\
 4121.327  & 27.0 &   -1.711  &    0.922  &  KUR  \\
 4121.329  & 27.0 &   -1.758  &    0.922  &  KUR  \\
 4121.329  & 27.0 &   -2.126  &    0.922  &  KUR  \\
 4121.331  & 27.0 &   -1.863  &    0.922  &  KUR  \\
 4121.332  & 27.0 &   -2.059  &    0.922  &  KUR  \\
 4121.336  & 27.0 &   -2.758  &    0.922  &  KUR  \\
 4121.336  & 27.0 &   -3.302  &    0.922  &  KUR  \\
 4121.337  & 27.0 &   -2.603  &    0.922  &  KUR  \\
 4121.338  & 27.0 &   -2.603  &    0.922  &  KUR  \\
 4121.338  & 27.0 &   -2.700  &    0.922  &  KUR  \\
 4121.338  & 27.0 &   -2.904  &    0.922  &  KUR  \\
\hline 
Ni& & & & \\
3858.297  &  28.0  &   -0.960  &    0.422  &  LAW  \\
5139.255  &  28.0  &   -1.097  &    3.655  &  SNE  \\
5476.904  &  28.0  &   -0.780  &    1.825  &  LAW  \\
\hline 
Sr& & & & \\
4077.709  &  38.1  &    0.167  &    0.000  &  SNE  \\
4215.519  &  38.1  &   -0.145  &    0.000  &  SNE  \\
\hline 
Y& & & & \\
4374.946  &  39.1  &    0.160  &    0.408  &  BIE  \\
\hline 
Ba& & & & \\
 4554.001  & 56.1 &   -0.636  &    0.000  &  GAL   \\
 4554.002  & 56.1 &   -1.033  &    0.000  &  GAL   \\
 4554.002  & 56.1 &   -0.636  &    0.000  &  GAL   \\
 4554.003  & 56.1 &   -0.636  &    0.000  &  GAL   \\
 4554.004  & 56.1 &   -1.033  &    0.000  &  GAL   \\
 4554.004  & 56.1 &   -0.636  &    0.000  &  GAL   \\
 4554.034  & 56.1 &    0.170  &    0.000  &  GAL   \\
 4554.034  & 56.1 &    0.170  &    0.000  &  GAL   \\
 4554.036  & 56.1 &    0.170  &    0.000  &  GAL   \\
 4554.050  & 56.1 &   -0.189  &    0.000  &  GAL   \\
 4554.053  & 56.1 &   -0.636  &    0.000  &  GAL   \\
 4554.053  & 56.1 &   -0.189  &    0.000  &  GAL   \\
 4554.054  & 56.1 &   -1.337  &    0.000  &  GAL   \\
 4554.056  & 56.1 &   -0.636  &    0.000  &  GAL   \\
 4554.057  & 56.1 &   -1.337  &    0.000  &  GAL   \\
 6141.725  & 56.1 &   -2.456  &    0.704  &  GAL   \\
 6141.725  & 56.1 &   -2.456  &    0.704  &  GAL   \\
 6141.727  & 56.1 &   -1.311  &    0.704  &  GAL   \\
 6141.727  & 56.1 &   -1.311  &    0.704  &  GAL   \\
 6141.728  & 56.1 &   -2.284  &    0.704  &  GAL   \\
 6141.728  & 56.1 &   -2.284  &    0.704  &  GAL   \\
 6141.729  & 56.1 &   -1.214  &    0.704  &  GAL   \\
 6141.729  & 56.1 &   -0.503  &    0.704  &  GAL   \\
 6141.729  & 56.1 &   -1.214  &    0.704  &  GAL   \\
 6141.729  & 56.1 &   -0.503  &    0.704  &  GAL   \\
 6141.730  & 56.1 &   -0.077  &    0.704  &  GAL   \\
 6141.730  & 56.1 &   -0.077  &    0.704  &  GAL   \\
 6141.730  & 56.1 &   -0.077  &    0.704  &  GAL   \\
 6141.731  & 56.1 &   -1.327  &    0.704  &  GAL   \\
 6141.731  & 56.1 &   -0.709  &    0.704  &  GAL   \\
 6141.731  & 56.1 &   -1.327  &    0.704  &  GAL   \\
 6141.731  & 56.1 &   -0.709  &    0.704  &  GAL   \\
 6141.732  & 56.1 &   -1.281  &    0.704  &  GAL   \\
 6141.732  & 56.1 &   -0.959  &    0.704  &  GAL   \\
 6141.732  & 56.1 &   -0.959  &    0.704  &  GAL   \\
 6141.733  & 56.1 &   -1.281  &    0.704  &  GAL   \\
 6496.898  & 56.1 &   -1.886  &    0.604  &  GAL   \\
 6496.899  & 56.1 &   -1.886  &    0.604  &  GAL   \\
 6496.901  & 56.1 &   -1.186  &    0.604  &  GAL   \\
 6496.902  & 56.1 &   -1.186  &    0.604  &  GAL   \\
 6496.906  & 56.1 &   -0.739  &    0.604  &  GAL   \\
 6496.906  & 56.1 &   -0.739  &    0.604  &  GAL   \\
 6496.910  & 56.1 &   -0.380  &    0.604  &  GAL   \\
 6496.910  & 56.1 &   -0.380  &    0.604  &  GAL   \\
 6496.910  & 56.1 &   -0.380  &    0.604  &  GAL   \\
 6496.916  & 56.1 &   -1.583  &    0.604  &  GAL   \\
 6496.916  & 56.1 &   -1.583  &    0.604  &  GAL   \\
 6496.917  & 56.1 &   -1.186  &    0.604  &  GAL   \\
 6496.918  & 56.1 &   -1.186  &    0.604  &  GAL   \\
 6496.920  & 56.1 &   -1.186  &    0.604  &  GAL   \\
 6496.922  & 56.1 &   -1.186  &    0.604  &  GAL   \\
\hline 
\hline
\end{longtable} 
\twocolumn
 \end{appendix}

\end{document}